\documentclass[published]{agujournal2019}

\usepackage{multibib}
\newcites{Supp}{References From the Supporting Information}

\usepackage{url} 
\usepackage[inline]{trackchanges} 
\usepackage{soul}
\journalname{JGR: Machine Learning and Computation}

\begin{document}

%
%


\title{A Next-Generation Snow Albedo Parameterization for Climate Modeling using Constrained Machine Learning}

%
%




\authors{A. Charbonneau\affil{1}, K. Deck\affil{1}, T. Schneider\affil{1}}

\affiliation{1}{California Institute of Technology}




\correspondingauthor{Andrew Charbonneau}{acharbon@caltech.edu}



\begin{keypoints}
\item Our data-driven albedo scheme achieves improved and generalizable predictions versus existing models at both fine and coarse scales
\item A simple scheme using only standard meteorological inputs provides competitive output at low computational burden
\item The utilized framework offers potential for efficient and scalable physical parameterizations in next-generation modeling regimes
\end{keypoints}

%
%

%
%


\begin{abstract}
We demonstrate a data-driven parameterization framework for snow albedo evolution using a constrained auto-regressive neural differential equation that directly predicts snow albedo change from standard meteorological inputs. After training with multi-year in-situ and satellite observations from a wide variety of locations, the scheme effectively reproduces daily albedo evolution across diverse climate zones, with median error under 7.5\% (RMSE $\approx$ 0.05), a 10-30\% improvement over established models. Furthermore, the model generalizes to sites not seen during training and scales from coarser grids to point locations. The scheme can easily incorporate new features as observational networks expand, offering an adaptive and computationally lightweight framework for next-generation land and climate models.
\end{abstract}

\section*{Plain Language Summary}
Snow reflects sunlight back to space, and small changes in its reflectivity (albedo) can influence our climate. Accurately modeling how snow albedo changes over time remains a challenge in climate models. We develop a data-driven approach that learns the rate of change of albedo using commonly available weather and snow measurements, instead of variables that would be unavailable or more uncertain within a climate model. The model is trained using several years of satellite and ground-based observations from a wide range of environments. Our model reproduces daily changes in snow albedo across many locations with lower error than widely-used alternatives, performing well for both ground measurements and satellite-scale data. Our approach can easily incorporate new observations and is computationally inexpensive, providing a flexible means for aiding next-generation earth science models and beyond. 

%
%

%


%
%
%
%

\section{Introduction}
Seasonal snowpacks play a crucial role in the cryosphere, influencing climate at small and large scales. They constitute an integrated response to a region’s weather, covering nearly half the Northern Hemisphere land surface in winter while also ranking among the fastest-changing land covers by season \cite{Dong_Ek_Hall_Peters-Lidard_Cosgrove_Miller_Riggs_Xia_2014, Klein_Stroeve_2002, Hall_Riggs_2007}. They play a significant role in terrestrial energy and water fluxes, with consequences for ecosystem and economic functions such as water supply, agriculture, and environmental hazards \cite{Lee_Gim_Park_2023, Hao_Bisht_Rittger_Stillinger_Bair_Gu_Leung_2023, Huang_Song_Yang_Yu_Liu_Che_Chen_Wu_Shu_Peng_et_al._2022}. However, snowpacks are sensitive to changing climates, and historical norms of accumulation and melt no longer reliably capture present-day patterns and timing \cite{Aguirre_Bozkurt_Sauter_Carrasco_Schneider_Jana_Casassa_2023, Rittger_Raleigh_Dozier_Hill_Lutz_Painter_2020}. Snow dynamics remain one of the most challenging aspects to represent in simulations of the water cycle \cite{Pirazzini_Leppanen_Picard_Lopez-Moreno_Marty_Macelloni_Kontu_Von_Lerber_Tanis_Schneebeli_et_al._2018}, emphasizing the need for improved snow models that can adapt and generalize to a range of possible futures at seasonal and multidecadal timescales.

Within snow models, one of the most critical parameters affecting the climate system is the snowpack albedo ($\alpha$), the ratio of reflected to incident shortwave radiation from the snow surface. Broadband snow albedo integrates wavelength-dependent snow albedo over the spectrum of incoming radiation \cite{Bair_Rittger_Skiles_Dozier_2019}:

\begin{linenomath*}
\begin{equation}
    \alpha = \frac{\int_{\lambda_1}^{\lambda_2}\alpha(\lambda)S(\lambda)d\lambda}{\int_{\lambda_1}^{\lambda_2}S(\lambda)d\lambda}
\end{equation}
\end{linenomath*}
where $\alpha(\lambda)$ is the snow's spectrally-resolved albedo, $S(\lambda)$ is the solar spectral irradiance at the snow surface, and $\lambda_1$ = 250 nm and $\lambda_2$ = 4000 nm are conventional cutoffs for the solar shortwave spectrum. Snow exhibits some of the highest natural albedos on Earth (0.7–0.9), making snow albedo and subsequent energy fluxes uniquely sensitive to perturbations or small errors. A small relative reduction in $\alpha$ creates a large relative increase in absorbed shortwave energy (1 – $\alpha$) which accelerates melt and lowers albedo further, amplifying a positive feedback on warming \cite{Bair_Rittger_Skiles_Dozier_2019} and leading to error propagation within global climate models (GCMs) \cite{Qu_Hall, Thackeray}. Through this mechanism, snow albedo exerts a strong control on the surface energy balance and overlying air circulation, and thus regional climate patterns and precipitation \cite{Hall_Riggs_2007, Malik_van_der_Velde_Vekerdy_Su_2014, Cordero_Sepulveda_Feron_Wang_Damiani_Fernandoy_Neshyba_Rowe_Asencio_Carrasco_et_al._2022, Kouki_Raisanen_Luojus_Luomaranta_Riihela_2022, Diro_Sushama_2018}. This influence propagates to global circulation, affecting processes such as ENSO, monsoons, and poleward retreat \cite{Wang_Schaaf_Chopping_Strahler_Wang_Roman_Rocha_Woodcock_Shuai_2012}, underscoring the need for accurate representation.

Reducing errors in modeled snow albedo is difficult given its variability and multiscale drivers. Snow albedo is governed by a suite of physical processes, including snow metamorphism, deposition of new snow or light-absorbing particles (LAPs) such as dust and black carbon, snow compaction and layer formation, and interaction with the evolving solar/atmospheric state (see \citeNP{Warren_1982, ZHANG_OHATA_2003}). However, snow albedo also varies strongly across short spatial and temporal scales due to topography and microclimate \cite{Molotch_Bales_2006, Williamson_Copland_Hik_2016}, and modification by LAPs often reflects both local and remote sources, decoupling changes from solely local snowpack processes \cite{Bair_Rittger_Skiles_Dozier_2019}. A main challenge in modeling snow albedo lies in representing the interplay of these processes across scales.

Most prevalent snow albedo models are built upon two formative ideas from the 1950s. The first was the application of basic scattering theory to model snow albedo analytically \cite{Dunkle_Bevans_1956}, with later extensions \cite{Wiscombe_Warren_1980} to incorporate multiple scattering, solar zenith angle, LAPs, and other snowpack properties. The second was the quantification of empirical surface-age and positive-temperature indices in an Army Corps of Engineers report \cite{Allard_1957} to estimate snow albedo decay. These ideas established the spectrum of theoretical versus empirical approaches to albedo modeling that are still in use today.

Results of these studies remain inconclusive, with complexity showing limited influence on overall parameterization skill and no consensus on which approach best suits large-scale modeling \cite{Gardner_Sharp_2010, Menard_Essery_Krinner_Arduini_Bartlett_Boone_Brutel-Vuilmet_Burke_Cuntz_Dai_et_al._2021, Essery_Morin_Lejeune_B_Menard_2013, Krinner_Derksen_Essery_Flanner_Hagemann_Clark_Hall_Rott_Brutel-Vuilmet_Kim_et_al._2018, Magnusson_Wever_Essery_Helbig_Winstral_Jonas_2015}. Simpler age-based schemes can reproduce global mean decay rates, but struggle to capture variability under changing conditions, routinely producing errors of 10–20\% and comparable biases outside their calibration domains \cite{Bair_Rittger_Skiles_Dozier_2019, Etchevers_Martin_Brown_Fierz_Lejeune_Bazile_Boone_Dai_Essery_Fernandez_et_al._2004, Damiani_Ishizaki_Feron_Cordero_2025, Abolafia_Rosenzweig_He_McKenzie_Skiles_Chen_Gochis_2022}. 
Complex or process-based models developed to address these limitations such as SNICAR, TARTES, and SNOWPACK \cite{Flanner_Arnheim_Cook_Dang_He_Huang_Singh_Skiles_Whicker_Zender_2021, Lehning_Bartelt_Brown_Fierz_2002, TARTES} demand input availability and process fidelity beyond what is contemporarily available at at GCM scales ($\sim10-100$km), leaving them uncertain or intractable even before accounting for increased computational cost 
\cite{Gardner_Sharp_2010, Damiani_Ishizaki_Feron_Cordero_2025, Essery_Morin_Lejeune_B_Menard_2013, Hao_Bisht_Rittger_Stillinger_Bair_Gu_Leung_2023, Bair_Rittger_Skiles_Dozier_2019, Menard_Essery_Krinner_Arduini_Bartlett_Boone_Brutel-Vuilmet_Burke_Cuntz_Dai_et_al._2021, Chan_2022, Gaillard_Vionnet_Lafaysse_Dumont_Ginoux_2025, Wang_Yang_Zhao_Zheng_Lu_Mamtimin_Ding_Li_Zhao_Li_et_al._2020, He_Flanner_Lawrence_Gu_2024}. Their sensitivity to poorly constrained inputs and site-specific calibration leave these and intermediate-complexity models with limited climate transferability and error ranges that frequently converge back to those of simpler schemes
\cite{Etchevers_Martin_Brown_Fierz_Lejeune_Bazile_Boone_Dai_Essery_Fernandez_et_al._2004, Lin_He_Abolafia-Rosenzweig_Chen_Wang_Barlage_Gochis_2025, Sarangi_Qian_Rittger_Bormann_Liu_Wang_Wan_Lin_Painter_2019, Oaida_Xue_Flanner_Skiles_De_Sales_Painter_2015, Shao_Xu_Li_Wang_Hao_2020, Lawrence_Oleson_Flanner_Thornton_Swenson_Lawrence_Zeng_Yang_Levis_Sakaguchi_et_al._2011}. 

Therefore, many GCMs continue to rely on combinations of simpler age-based parameterizations for tractability, despite wide performance spreads highly sensitive to model choice, calibration sites, and validation approaches, particularly during the ablative period \cite{Etchevers_Martin_Brown_Fierz_Lejeune_Bazile_Boone_Dai_Essery_Fernandez_et_al._2004, Essery_Morin_Lejeune_B_Menard_2013, Magnusson_Wever_Essery_Helbig_Winstral_Jonas_2015, Gunther_Marke_Essery_Strasser_2019}. Snow albedo parameterizations exert larger influence on snow simulations than those for other snow processes due to variability in albedo decay rates affecting snowpack decay, which can alter snow season length by 25–50+ days \cite{Lee_Gim_Park_2023, Wang_Schaaf_Chopping_Strahler_Wang_Roman_Rocha_Woodcock_Shuai_2012, Etchevers_Martin_Brown_Fierz_Lejeune_Bazile_Boone_Dai_Essery_Fernandez_et_al._2004, Essery_Morin_Lejeune_B_Menard_2013, Abolafia_Rosenzweig_He_McKenzie_Skiles_Chen_Gochis_2022}. As GCM resolutions continue to increase toward kilometer or sub-kilometer scales \cite{Aguirre_Bozkurt_Sauter_Carrasco_Schneider_Jana_Casassa_2023, Gaillard_Vionnet_Lafaysse_Dumont_Ginoux_2025, Voordendag_Reveillet_MacDonell_Lhermitte_2021, Ye_Cheng_Hao_Yu_Ma_Liang_Shen_2023, Ding_Liang_Ma_He_Jia_Wang_2024}, issues of generalization, input sensitivity, and computational burden will only become more acute. This highlights the need for computationally efficient but accurate and generalizable schemes for the next generation of GCMs, which are more data-informed than previous generations in an effort to reduce recalcitrant errors (for example, in surface energy fluxes) yet still face similar input uncertainty of desired variables like snow grain size or LAP concentration. This necessitates a framework that can represent diverse albedo decay behaviors across climates while still remaining computationally efficient, transferable, and dependent only on widely available inputs.

Recent computational advances and data volume have inspired new albedo parameterization efforts emphasizing grain size, LAP concentration, and spectral resolution such as \citeA{Bair_Rittger_Skiles_Dozier_2019}, \citeA{Gardner_Sharp_2010}, \citeA{Saito_Yang_Loeb_Kato_2019}, and \citeA{Strom_Svensson_Honkanen_Asmi_Dkhar_Tayal_Sharma_Hooda_Meinander_Lepparanta_et_al._2022}. These studies report accurate performance across snow conditions but still inherit the input sensitivity of their predecessors, limiting their utility in GCMs. Machine Learning (ML) instead offers potential, as ML approaches can capture nonlinear interactions directly from data while avoiding the systematic bias inherent to presupposing explicit functional forms \cite{Damiani_Ishizaki_Feron_Cordero_2025, Demil_Haghighi_Klove_Oussalah_2025, Fleming_Rittger_Oaida_Taglialatela_Graczyk_2024}. However, efforts specific to snow albedo modeling remain scarce. Many focus on snow coverage to improve data fidelity, or largely post-process input reflectance products or predetermined albedo, lacking the physical basis for incorporation into GCMs where such inputs are unavailable \cite{Ding_Liang_Ma_He_Jia_Wang_2024, Zhao_Ding_Wang_Qu_2025, Chen_Xiao_Zhang_Pellikka_Liu_Liu_2025}. These efforts likewise find limited transferability beyond calibration sites, with many errors still plateauing near $\sim$10\% \cite{Ding_Liang_Ma_He_Jia_Wang_2024, Demil_Haghighi_Klove_Oussalah_2025, Kiem_Hammerle_Montagnani_Wohlfahrt_2024}, and many ML-based approaches do not inherently respect physical constraints \cite{De_Michele_Avanzi_Ghezzi_Jommi_2013, Gao_Zhang_Shen_Zhang_Ai_Zhang_2021}.

This work presents a novel parameterization, which uses the machine learning framework developed in \citeA{Charbonneau_Deck_Schneider_2025} to predict the time derivative of the snow albedo given only meteorological inputs and the present snow albedo as inputs. The predictive model output, which obeys prescribed constraints, is then integrated over time to obtain the broadband snow albedo. It is able to outperform a suite of established models over a large variety of sites from only a single calibration on coarse-scale data, with results that extend from the coarser scales of present data products to fine-scale point validation without recalibration. We use the U.S. Snow Telemetry (SNOTEL) network to gather site-level snow state data, in combination with the STC/MODSCAG-MODDRFS 500-m albedo product \cite{RITTGER_Lenard_Palomaki_Brodzik_Stillinger_Bair_Dozier_Painter_2024} to create coarse-level data, along with established validation site data and automated weather station data to allow testing over a variety of climates. The approach is also accessible to many computational systems and fast to train and prototype (under one minute for training on a single CPU, and able to process millions of inputs per second). It is adaptable as new variables (e.g., characterizing aerosol deposition) become available, demonstrating versatile capability for advancing climate modeling.

\section{Methodology}

We utilize a ML framework developed in 2022 and introduced in \citeA{Charbonneau_Deck_Schneider_2025} to construct our parameterization, which enforces prescribed constraints (supplied analytical functions, additional parameterizations, etc.) using fixed layers during both training and operation, enabling constrained learning without penalizing the loss function. This scheme was shown to enable efficient, physically consistent, and generalizable parameterizations from simple predictive structures, applicable to a variety of systems and capable of performing on par with more complex models. We combine remotely sensed data in addition to point observations to support benchmarking across scales and a variety of climates. By prioritizing observational data rather than age-based or assimilated features, we avoid circular bias from reanalysis-driven inputs, break from continued reliance on age indices, and achieve more faithful representations of underlying snow processes for usage within GCMs.

\subsection{Model}

We model the rate of change in snow albedo with a simple neural network, $M$ driving an ordinary differential equation for the evolution of snow albedo:
\begin{figure}[ht!]
\centerline{\includegraphics[width = 25pc]{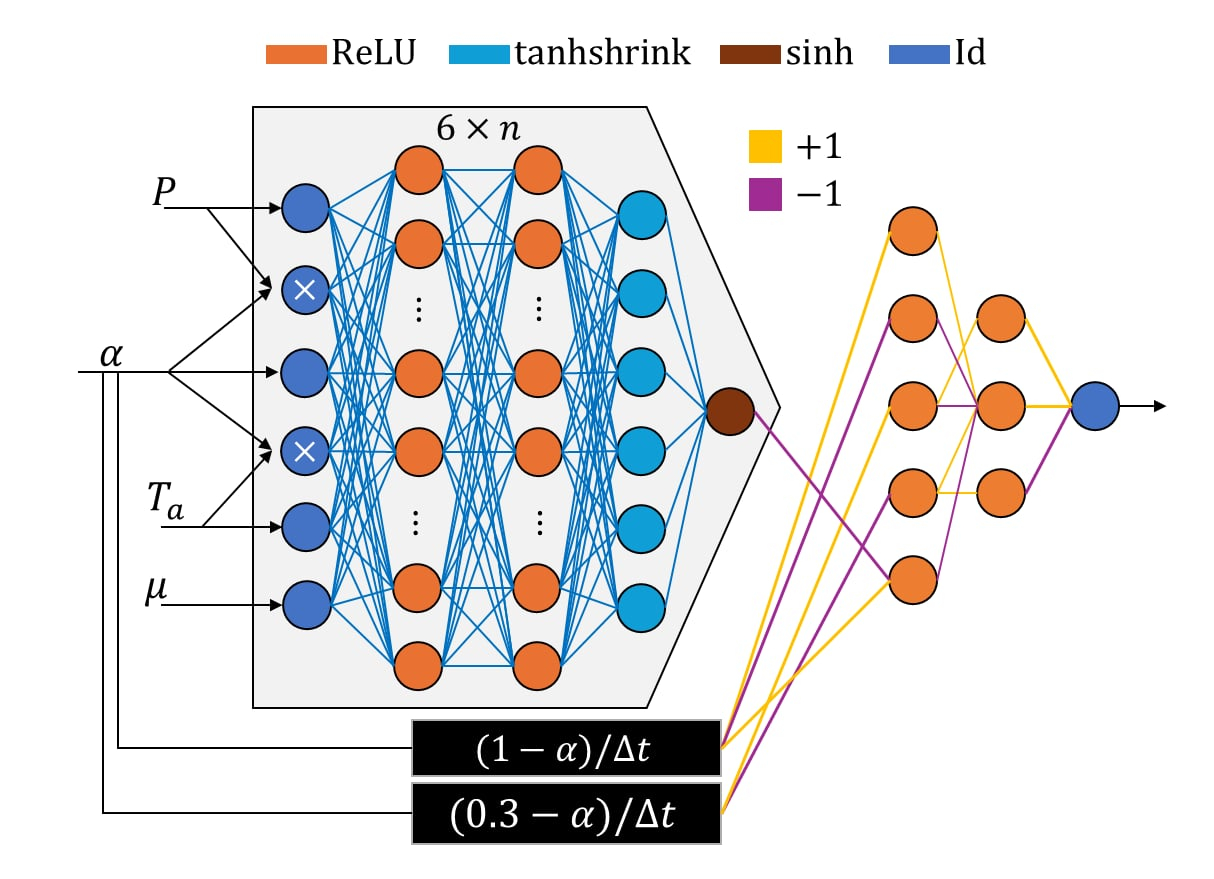}}
\caption{\footnotesize \textbf{Structure of $M$}. The predictive portion of the parameterization consists of a simple 3-layer dense neural network with an aggregation to a final output. Nodes are colored by their aggregation function, trainable weights (and a bias term) are shown in blue, and fixed weights (no bias term) are shown in yellow (+1) or purple (-1). One hyperparameter $n$ determines the layer widths. The white ``x" indicates multiplication to allow usage of the cross-term inputs $\alpha T_a$ and $\alpha P$. A threshold constraint layer (see Appendix A in \citeNP{Charbonneau_Deck_Schneider_2025}) is included around the predictive component to keep the accumulated albedo between 0.3 and 1 for time-step $\Delta t$, even during training.}
\label{f:structure}
\end{figure}

\begin{linenomath*}
\begin{equation} \label{eq:diffeq}
    \frac{d\alpha}{dt}=M\left(\alpha,\ T_a,\ P,\ \mu\right)
\end{equation}
\end{linenomath*}
where $T_a$ is air temperature ($^\circ$C), $P$ is liquid water-equivalent snowfall rate (m s$^{-1}$), $\mu$ is the cosine of the solar zenith angle (at solar noon for daily data), and $\alpha$ is current albedo (0–1). The network structure is shown in Fig. \ref{f:structure}, and constraint functions are picked only to limit the subsequently updated snow albedo value between an applicable range of 0.3 and 1 \cite{Henneman_Stefan_1999}, which restricts reflectivities to the physical range [0, 1]. The constraints are enacted by fixed layers that remain during both training and inference, guaranteeing compliant outputs by construction (prohibiting violative outputs that would arise from their removal) and improving gradient updates within valid bounds during training. This eliminates the need to add loss function penalties, which can impede optimization of the primary objective \cite{Rahimi:2023}, and was shown to protect against overfitting and enable learning of more extrapolative final models, compared to identical models trained without constraints that were only applied afterwards \cite{Charbonneau_Deck_Schneider_2025}.

Unlike our previous work and other snow albedo schemes, we did not want to \textit{a priori} restrict albedo increases to occur only under precipitation with the selected constraints, since phenomena like the smoothing of the snow surface during melt or overnight freezing can increase the albedo absent new precipitation. Our utilized boundaries relax this prevalent but narrow assumption. We also chose not to impose additional structure in this initial study --- such as preventing albedo increases under higher air temperatures, or preventing albedo decreases under sufficient precipitation --- instead favoring less restrictive bounds which maximally limit the network (not guide the network internal behavior), though the framework can readily represent such choices. We note that while prohibiting albedos greater than 1 or negative albedos imposes physical limits on the generated dynamics, the actual lower value of 0.3 is empirically motivated. This could introduce systematic biases for more extreme snow environments where darker albedos are needed but are architecturally prohibited by the model. However, alteration of this bound is straightforward and does not inherently require retraining of the model weights.

Choosing initial boundaries solely to maintain expected global limits and physical bounds on reflectivity did not presuppose any assumptions about albedo dynamics. This choice allowed us to investigate the extent to which physically consistent relationships could be learned from the data without imposing additional model structure, or whether specifying additional behavior might be necessary. The prognostic format mirrors the differential equations used for other GCM variables, aiding integration, and reflects findings that prognostic parameterizations are more consistent than other forms \cite{Essery_Morin_Lejeune_B_Menard_2013}. Learning the rate or change directly via a neural network permits a nonlinear expression of a diverse range of rates from an otherwise simplistic input space, addressing one of the main limitations in previous simple or prognostic schemes \cite{Etchevers_Martin_Brown_Fierz_Lejeune_Bazile_Boone_Dai_Essery_Fernandez_et_al._2004, Abolafia_Rosenzweig_He_McKenzie_Skiles_Chen_Gochis_2022} and improving representative capacity to observed variability.

We deliberately employ a small black-box neural network (NN) as the internal predictive model within the constraint structure, rather than linear regression, tree-based methods, or other ML approaches. For training volumes available in climate parameterizations and this study, reported tabular-data advantages of tree ensembles diminish \cite{grinsztajn2022treebasedmodelsoutperformdeep, gorishniy2023revisitingdeeplearningmodels}, while NNs exhibit stronger expressive capacity for the continuous nonlinear interactions governing albedo. While white-box models like single trees are interpretable, ensemble methods lose most structural transparency, offering little practical interpretability over an intentionally compact NN for GCM integration. Explicit cross-terms and ReLU in the first layer reduce the required depth/width to otherwise learn multiplication and enable explicit representation of ``controlled" behaviors (e.g., $(\alpha-\alpha_0)*H(P-P_0)$; $H$ is the Heaviside function) with fewer parameters and operations. Unlike trees, inference with a NN is continuous, fully differentiable, more GPU/cache-scalable without conditional branching, and more compatible with Kalman/gradient-based calibration workflows utilized in GCM sub-model tuning. Our framework with NNs was also shown to permit physically generalizable extrapolation beyond the convex hull of the training targets \cite{Charbonneau_Deck_Schneider_2025}, whereas tree models cannot, which is important in evolving climate systems where no data is wholly representative. Convolutional or spatiotemporal networks were not pursued, as the governing processes are effectively localized in the selected input features at GCM-relevant scales, and spatial aggregation amplifies missing-data and uncertainty propagation. We emphasize these choices showcase an informed baseline rather than a final architecture, as the framework remains extensible to alternative model classes and other constraint choices.  

Since model complexity shows little correlation with model skill, using a simple network aids computational efficiency, robustness to input uncertainty, and protection against overfitting; all necessary for high-resolution GCMs \cite{Essery_Morin_Lejeune_B_Menard_2013, Magnusson_Wever_Essery_Helbig_Winstral_Jonas_2015}. Grain size is deliberately excluded; although it often improves diagnostic accuracy, it increases sensitivity and couples performance to an uncertain or unavailable input at GCM scales, and measurements transfer poorly across regions, risking bias and reduced generalizability. The utilized variable set also breaks from a continued reliance on snow age. Using a 1D model over well-characterized local state variables (i.e., not directly ingesting/requiring spatiotemporal information) allows broad applicability over any spatial grid and better representation of inherently universal processes, while the lack of explicit time dependence within the predictive component (only time-step dependence in the constraints) ensures the scheme can be run at any GCM time step without retraining \cite{Charbonneau_Deck_Schneider_2025}.

The proposed architecture should be able to emulate physically plausible relationships present in the training data, but this chosen construction does not explicitly enforce conservation of energy, radiative transfer equations, snow microstructure evolution, or other mechanistic constraints that can shape the dynamics between the bounds. Physically realistic behavior only arises from the combination of bounds and training data relationships rather than from guaranteed adherence to snow radiative transfer constraints. This model should augment and interact with but not replace radiative-transfer schemes, which resolves physics unavailable to the present parameterization and serves different purposes within a larger model. Rather, this scheme provides an empirical representation of the snow albedo emerging from processes that would otherwise require substantially more complex, uncertain, or unavailable intermediate state variables, and can be used within a larger model to characterize the reflectivity which is needed in determining radiative transfer.

\subsection{Data}

We combined in-situ observations from the U.S. Snow Telemetry (SNOTEL) network with the remotely sensed STC/MODSCAG-MODDRFS 500-m albedo product (hereafter ``SMM"). SNOTEL provides collocated ground measurements of snow and meteorological variables. In recent years, over 20 SNOTEL sites have also begun to collect surface radiation data, enabling direct snow albedo derivation for site-level validation. For earlier periods, snow albedo timeseries were extracted from the corresponding SMM grid point. SMM improves on earlier MODIS-based products through enhanced corrections for clouds, canopy viewing angle, and ground slope, reducing errors by nearly threefold against widely used products (e.g., MOD10A1), especially in forested regions where most SNOTEL sensors are located. It more accurately represents snow albedo instead of mixed surface albedo as well as snow onset, disappearance, and melt season dynamics down to lower snow cover fractions \cite{Hao_Bisht_Rittger_Stillinger_Bair_Gu_Leung_2023, Aguirre_Bozkurt_Sauter_Carrasco_Schneider_Jana_Casassa_2023, Rittger_Raleigh_Dozier_Hill_Lutz_Painter_2020, Sarangi_Qian_Rittger_Bormann_Liu_Wang_Wan_Lin_Painter_2019, Rittger_Painter_Dozier_2013, Stillinger_Rittger_Raleigh_Michell_Davis_Bair_2023}, with typical validation errors of 0.05–0.08 \cite{Dong_Ek_Hall_Peters-Lidard_Cosgrove_Miller_Riggs_Xia_2014, Bair_Rittger_Skiles_Dozier_2019, Painter_Rittger_McKenzie_Slaughter_Davis_Dozier_2009, Williamson_Copland_Hik_2016, Painter_Bryant_Skiles_2012, Hatchett_Koshkin_Guirguis_Rittger_Nolin_Heggli_Rhoades_East_Siirila_Woodburn_Brandt_et_al._2023, Palomaki_Rittger_Lenard_Bair_Dozier_Skiles_Painter_2025}. Usage of SMM has yielded 10–25\% gains in hydrological modeling skill, especially during melt season, underscoring its suitability for next-generation models \cite{Fleming_Rittger_Oaida_Taglialatela_Graczyk_2024}. We are additionally not aware of prior studies using the new SNOTEL albedo data for model validation, offering a novel benchmark. To further test generalization and provide comparable benchmarks, we incorporated standard evaluation sites from Col de Porte \cite{Lejeune_Dumont_Panel_Lafaysse_Lapalus_Le_Gac_Lesaffre_Morin_2019}, K{\"u}htai \cite{Krajci_Kirnbauer_Parajka_Schoer_Bloschl_2017}, and Mammoth Mountain \cite{Bair_Davis_Dozier_2018}, and include additional data from Sodankylä \cite{Essery_Kontu_Lemmetyinen_Dumont_Menard_2016}, the Upper Rofental \cite{Warscher_Marke_Rottler_Strasser_2024}, the Himalayas \cite{Stigter_Steiner_Koch_Saloranta_Kirkham_Immerzeel_2021, Shea_Wagnon_Immerzeel_Biron_Brun_Pellicciotti_2015}, and one site in Chile from the Centro de Estudios Avanzados en Zonas Áridas (CEAZA) meteorological network, to test performance across out-of-sample elevations and climates.

Both SMM and site sensor data have known quality issues. For SMM, accuracy varies by location and with vegetation cover, and it can exhibit geolocation errors of up to one pixel \cite{Rittger_Raleigh_Dozier_Hill_Lutz_Painter_2020, Bair_Rittger_Skiles_Dozier_2019, Demil_Haghighi_Klove_Oussalah_2025, Bair_Dozier_Stern_LeWinter_Rittger_Savagian_Stillinger_Davis_2022}. SMM’s underlying calibration data is not inherently representative of all climates \cite{Keuris_Hetzenecker_Nagler_Molg_Schwaizer_2023} and can influence poor distinction between snow types \cite{Hao_Bisht_Rittger_Stillinger_Bair_Gu_Leung_2023}, which can bias snow cover fractions and subsequently derived albedos (\citeNP{Lee_Gim_Park_2023}; see Appendix A). Systematic underestimation of snow-covered areas, summertime values, and misrepresentation of shallow snowpacks ($<$5 cm) suggest non-snow albedo continues to contaminate reported values \cite{Ding_Liang_Ma_He_Jia_Wang_2024, Palomaki_Rittger_Lenard_Bair_Dozier_Skiles_Painter_2025, Klein_2003}. For SNOTEL and other site sensors, precipitation is often underreported, and sensor records can contain biased or unphysical values \cite{Meyer_Jin_Wang_2012, Hill_Burakowski_Crumley_Keon_Hu_Arendt_Wikstrom_Jones_Wolken_2019, Warscher_Marke_Rottler_Strasser_2024}. Because no universal cleaning methods exist—particularly for the new SNOTEL albedo records—we developed a custom procedure (see Appendix A) to quality-control, de-noise, smooth, and scale albedo timeseries, supplemented with established methods for other variables \cite{Serreze_Clark_Armstrong_McGinnis_Pulwarty_1999, Yan_Sun_Wigmosta_Skaggs_Hou_Leung_2018, Livneh_Deems_Schneider_Barsugli_Molotch_2014, atwood2023evaluation}. From cleaned records, $d\alpha / dt$ was derived for days with complete data (excluding $\Delta t >$ 1), creating a table of input feature values and an associated target $d\alpha / dt$ that could be shuffled for training. Daily zenith angle cosine was computed with \texttt{Insolations.jl} \cite{Insolation.jl}, and snow–rain fractions were derived using the parameterization from \cite{Jennings_Winchell_Livneh_Molotch_2018}, shown to be over 85\% accurate. Data were scaled but not normalized (no shift) to preserve the physical meaning of input zeroes. Data scaling constants (standard deviations for inputs, absolute maxima for targets) were embedded within the neural network to ensure consistent preprocessing and postprocessing, eliminating the need for users to manually apply input and output scaling during model evaluation. This yielded roughly 1.6 million site-days with SMM albedo and 15,300 with direct observations.

\subsection{Model Training}

The combination of SNOTEL sites with the SMM values along with the other validation sites provides two albedo spatial scales for model evaluation: point-scale timeseries from ground sensors to assess emulation of underlying albedo processes, and extracted 500 m grid-point timeseries to gauge representation of coarse-grained dynamics more relevant for climate models and evolving GCM needs. To emphasize generalizability, the model was trained on a small subset of the coarser SMM data (40 locations out of 740; see below), saving the rest for coarse-scale testing. We also used the same resulting calibration to test over out-of-sample ground-level records from varying climate types, forcing a more stringent test of the parameterization’s capacity to extract and learn relationships that should transfer across scales. We used only the daily data from the 40 SNOTEL sites overlapping with our previous study for training, as these were shown to exhibit sufficient diversity in elevation and local climate to enable robust parameterization (see spread in altitude and snowpack types in Fig. 3 from \citeNP{Charbonneau_Deck_Schneider_2025}, as well as Fig. S1 in the Supporting Information for all sites in this study). This left 700 unused SNOTEL sites for coarser testing and 31 site-level test locations, providing extensive performance statistics while highlighting the framework’s ability to generalize from a relatively smaller training sample ($\sim$87000 batched and shuffled site-day samples from SMM). The model was trained with $L_2$ loss, using $n$ = 2 for the width hyperparameter (see Appendix C for hyperparameter selection), creating a total of 331 weights. Training for 100 epochs (100 passes over all training data) with batch sizes of 256 requires $<$30 s on a single Intel i9 CPU, with a storage footprint under 2 kB. This likewise competes with the usual training time advantage associated with tree ensembles and other non-adjoint machine learning approaches. Training was implemented in Julia’s \texttt{Flux} framework \cite{Innes_2018} using the root-mean-square propagation (RMSProp) optimizer \cite{Hinton:2012}. Data sources and variables are summarized in Table \ref{t:varsummary}.

\begin{table}[ht!]
\centering
\caption{\footnotesize \textbf{Summary of data and variables.} This table denotes the names of utilized data sources and their purpose, as well as the variables mentioned throughout the paper. The Days column refers to the number of sensor days with complete records of at least $\alpha, P$, and $T_a$. ``Water equivalent" is abbreviated as ``w. eq". Bulk Density is represented as a unitless fraction of the density of liquid water, 1000 kg m$^3$. Only variables listed in Eq. \ref{eq:diffeq} are used within $M$. The spread of snowpack types and geographic locations of all sites can be found in Fig. S1 of the Supporting Information.}
\label{t:varsummary}
\resizebox{\textwidth}{!}{%
\begin{tabular}{lcccc|cccl}
\hline
\multicolumn{5}{c|}{Data} & \multicolumn{4}{c}{Variables} \\
Source & Sites & Days & Scale & Purpose & Symbol & Units & Name & \multicolumn{1}{c}{Description} \\ \hline
SNOTEL/SMM Albedo & 40 & 86841 & 500m & Training & $\alpha$ & None & Snow Albedo & Derived, see Appendix A \\
SNOTEL/SMM Albedo & 700 & 1527868 & 500m & Testing & $P$ & ms$^{-1}$ & Snowfall (w.eq.) & Direct or derived (Appendix A) \\
SNOTEL/SNOTEL Radiation & 23 & 5318 & Point & Testing & $T_a$ & $^\circ$C & Air Temp. & Averaged to daily level \\
Col De Porte, France & 1 & 2340 & Point & Testing & $\mu$ & None & Zenith Angle & Cosine at solar noon \\
K{\"u}htai, Austria & 1 & 4532 & Point & Testing & $z$ & m & Snow depth & Direct data \\
Sodankyla, Finland & 1 & 317 & Point & Testing & SWE & m & Snow w.eq. & Direct data \\
Bella Vista, Upper Rofental & 1 & 457 & Point & Testing & $\rho$ & None & Bulk Density & Derived as SWE/$z$ \\
Proviantdepot, Upper Rofental & 1 & 741 & Point & Testing & $\tau$ & days & Snow Age & See below \& Appendix A \\
Yala Basecamp, Himalayas & 1 & 126 & Point & Testing & $\alpha_g$ & None & Ground Albedo & Derived, see Appendix A \\
El Tapado, Chile (CEAZA Network) & 1 & 268 & Point & Testing & DOY & None & Day of Year & Derived from Date \\
Dozier Snow Study Site (CUES) & 1 & 1490 & Point & Testing & Date & Date & Date & Direct data \\
\textbf{TOTAL} & \textbf{771} & \textbf{1630298} &  &  & $T_s$ & $^\circ$C & Surface Temp. & Estimated, $\min{(0, T_a)}$ \\ \hline
\end{tabular}%
}
\end{table}

\subsection{Model Testing and Comparison}

\citeA{Magnusson_Wever_Essery_Helbig_Winstral_Jonas_2015} notes new parameterizations are often benchmarked without comparison against existing models, limiting assessment of their true value. To address this, we compare our parameterization against a comprehensive set of alternatives (Table \ref{t:models}), representative of those used in current GCMs \cite{Lee_Gim_Park_2023} or proposed in prior studies. Consistent with \citeA{Magnusson_Wever_Essery_Helbig_Winstral_Jonas_2015} and similar efforts, alternative schemes are primarily compared as-published beyond adjusting time-step-dependent parameters to daily resolution, since many have already been tuned for global use. Additional recalibration may reduce absolute errors but rarely alters the rough ranking or performance categories of schemes \cite{Essery_Morin_Lejeune_B_Menard_2013}, making such adjustments unnecessary. However, since our neural parameterization must be calibrated to be compared, for fair and holistic comparison a supplemental benchmarking was included for all schemes upon recalibrating them to the same data (see the Supplementary Information for equations and parameter values).

\begin{sidewaystable}
\caption{\footnotesize \textbf{Parameterizations compared in this study.} We benchmark the proposed parameterization against an exhaustive (and fundamental, as others are similar to those below) list of other simple forms (no grain size, only available macroscopic state variables) used in past and present GCMs or proposed in literature, noting their original appearance and reference consulted for parameter values. The table denotes each scheme’s symbol for usage throughout the paper, as well as the model type (prognostic-rate, prognostic-value, or diagnostic, as PR, PV, and D, respectively), whether the scheme uses a continually- (C) or threshold- (T) based reset of internal snow age/indices, and whether the scheme was calibrated autoregressively (AR) or via regression (yes[Y]/no[N], see below). Equations are provided in the Supporting Information. Models explicitly found in CMIP6 \cite{gmd-9-1937-2016} are denoted with an asterisk. Some other CMIP6 parameterizations are very similar to listed table members (e.g. UTOPIA, BATS, $T_a$ Interpolation), or constant values. Other listed schemes have also been used in non-CMIP6 global or regional modeling.}
\label{t:models}
\resizebox{\textwidth}{!}{%
\begin{tabular}{lcccccccl}
\hline
\multicolumn{1}{c}{Name} & Symbol & Identifiable Origins & Referenced Paper & Variables & Type & $P$ Reset & AR & \multicolumn{1}{c}{Has Been In:} \\ \hline
ACE Curve, Polynomial & $AC$ & \citeA{Allard_1957} & \citeA{Molotch_Bales_2006} & $\tau$ & D & T & N &  \\
Exponential Decay (Snow Age) & $ED$ & \citeA{riley1969snowmelt} & \citeA{Kondo_Yamazaki_1990} & $\tau$ & D & T & N & US ACE HEC-HMS \\
Exponential Decay ($T_a$ Index) & $ET$ & \citeA{ranzi1991physically} & \citeA{ranzi1991physically} & $P, T_a$ & D & T & Y &  \\
Logistic/Exponential Decay (Ta Index) & $LE$ & \citeA{Pike_Riley_Toal_Burnham_2024} & \citeA{Pike_Riley_Toal_Burnham_2024} & $P, T_a, z$ & D & T & Y & - \\
VAS & $VA$ & \citeA{Malik_van_der_Velde_Vekerdy_Su_2014} & \citeA{Malik_van_der_Velde_Vekerdy_Su_2014} & $P, T_a, z, \alpha_g$ & D & T & Y & - \\
Brock 2000 & $BR$ & \citeA{Brock_Willis_Sharp_2000} & \citeA{Brock_Willis_Sharp_2000} & $P, T_a, \mathrm{SWE}, \alpha_g$ & D & T & Y & - \\
Marshall (``Simple" Depth) & $MD$ & \citeA{marshall1989physical} & \citeA{marshall1989physical} & $\tau, \mathrm{SWE}, \alpha_g$ & D & T & N & - \\
Oerlemans \& Knap & $OK$ & \citeA{Oerlemans_Knap_1998} & \citeA{Liu_Ma_Menenti_Su_Yao_Ma_2021} & $\tau, z, \alpha_g$ & D & T & N & Noah \\
ESCIMO & $EC$ & \citeA{Rohrer_Mario_Bruno_1991} & \citeA{Strasser_Marke_2010} & $\tau, T_a$ & D & T & N & ESCIMO \\
Interpolation (Ta)* & $IT$ & \citeA{Roeckner_Arpe_Bengtsson_Brinkop_Dumenil_Esch_Kirk_Lunkeit_Ponater_Rockel_et_al._1992} & \citeA{Molders_Luijting_Sassen_2007} & $T_a$ & D & - & N & ECHAM, ECMWF, JSBACH3 \\
Interpolation (Density) & $ID$ & \citeA{GREUELL_KONZELMANN_1994} & \citeA{GREUELL_KONZELMANN_1994} & $\rho$ & D & - & N & SOMARS, RACMO2, CLSM \\
Noah Exponential & $LV$ & \citeA{Anderson_1968} & \citeA{Livneh_Xia_Mitchell_Ek_Lettenmaier_2010} & $\tau$, DOY & D & T & N & Noah, DHSVM, VIC \\
SNOWPACK-like & $SP$ & \citeA{Lehning_Bartelt_Brown_Fierz_2002} & This & $P, T_a, \mu, z, \rho$ & D & - & N & SNOWPACK \\
Zhongyuan \& Ohata Density & $ZO$ & \citeA{1989-fg} & \citeA{ZHANG_OHATA_2003} & $\rho$ & D & - & N & - \\
NASA GISS* & $NG$ & \citeA{Hansen_Russell_Rind_Stone_Lacis_Lebedeff_Ruedy_Travis_1983} & \citeA{Hansen_Russell_Rind_Stone_Lacis_Lebedeff_Ruedy_Travis_1983} & $P$ & D & C & Y & GISS Model II, ORCHIDEE, CLSM \\
CLASS* & $CL$ & \citeA{Verseghy_1991} & \citeA{Li_Zhang_Wang_Liu_Ju_Mamtimin_2023} & $P, \alpha$ & PV & C & Y & Noah-MP, ISBA, CLASS \\
UTOPIA & $UT$ & \citeA{Verseghy_1991}, \citeA{Baker_Ruschy_Wall_1990} & \citeA{Cassardo_2015} & $P, \alpha, T_s, \mu$ & PV & C & Y & UTOPIA \\
HTESSEL* & $HT$ & \citeA{Verseghy_1991}, \citeA{Baker_Ruschy_Wall_1990} & \citeA{Dutra_Balsamo_Viterbo_Miranda_Beljaars_Schar_Elder_2010} &$\alpha, P, \mathrm{SWE}, T_s$ & PV & C & Y & HTESSEL/ECMWF \\
Giddings and La Chapelle & $GC$ & \citeA{Giddings_LaChapelle_1961} & \citeA{o1972solar} & $z, \alpha_g$ & D & - & N & - \\
JULES* & $JL$ & \citeA{essery2001moses} & \citeA{Best_Pryor_Clark_Rooney_Essery_Menard_Edwards_Hendry_Porson_Gedney_et_al._2011} & $P, T_s$ & D & C & Y & JULES \\
Marks 1988 & $MS$ & \citeA{marks1988climate} & \citeA{marks1988climate} & $\tau, \mu$ & D & T & N & SNOBAL \\
BATS* & $BA$ & \citeA{Dickinson_Henderson-Sellers_Kennedy_Wilson_1986} & \citeA{Abolafia_Rosenzweig_He_McKenzie_Skiles_Chen_Gochis_2022} & $P, T_s, \mu$ & D & C & Y & Many (Noah-MP, WRF CoLM...) \\
ISBA* & $IS$ & \citeA{Brun_David_Sudul_Brunot_1992} & \citeA{Decharme_Brun_Boone_Delire_Le_Moigne_Morin_2016} & $z, T_s, T_a, P$ & D & C & Y & Crocus, ISBA MAR \\
Neural Network & $NN$ & Prev. Work \cite{Charbonneau_Deck_Schneider_2025} & This & $\alpha, P, T_a, \mu$ & PR & - & N & - \\ \hline
\end{tabular}}
\end{sidewaystable}

The benchmarked schemes fall into diagnostic or prognostic classes, with some incorporating internal snow-age or other ``latent-state" formulations which rendered them unable to be trained via regression like the neural scheme (see below). All schemes except ours predict albedo directly rather than its derivative. This distinction leads to three equations for generating time-stepped outputs, diagnostic (Eq. \ref{eq:diagnostic}), prognostic-value (Eq. \ref{eq:prognostic}), and prognostic-rate (Eq. \ref{eq:nn_timestep}, our neural parameterization):
\begin{linenomath*}
\begin{equation} \label{eq:diagnostic}
\widehat{\alpha_i}=f\left(\theta_i\right)
\end{equation}
\end{linenomath*}
\begin{linenomath*}
\begin{equation} \label{eq:prognostic}
\widehat{\alpha_i}=f\left(\widehat{\alpha_{i-1}},\ \theta_i\right)
\end{equation}
\end{linenomath*}
\begin{linenomath*} 
\begin{equation} \label{eq:nn_timestep}
\widehat{\alpha_{i+K}}=\alpha_i+\left(K\Delta t\right)M\left(\widehat{\alpha_i},\ P_i,\ T_{a,i},\ \mu_i\right)
\end{equation}
\end{linenomath*}
where a hat indicates the model prediction (and no hat for the observation) and $\theta_i$ indicate the scheme’s (denoted $f$) relevant set of non-albedo input variables at step $i$. All schemes were initialized to observed values $\hat{\alpha_1} = \alpha_1$. For the neural scheme, the integer $K$ denotes the step size across sequential days ($K$=1) or ``gaps" from missing/cleaned/no-snow periods ($K>$1). We note that when traversing data gaps via an Euler step with $K>1$, the resulting albedo could violate the bounds dictated by the predictive model (designed for step-size K=1), though such error should be attributed to the time-stepping choice, not the parameterization itself. To mitigate this, following our previous work \cite{Charbonneau_Deck_Schneider_2025}, a maximum gap size of $K_\mathrm{max} = 5$ days was permitted before timeseries were ``reset" to observed values $\widehat{\alpha_{i+K}} = \alpha_{i+K}$ if $K > K_\mathrm{max}$, and unphysical albedos from $1 < K < K_\mathrm{max}$ traversals were likewise clamped to [0.3,1] to avoid attributing error from the time-stepping choice to the parameterization for fair evaluation, as such issues would not occur within a global model setting. Other schemes were reset for any gaps, giving them a ``best-case" comparison. Latent-state schemes were not resumed upon reaching a gap until receiving at least $Q_0 = 3$ mm of liquid-equivalent new snowfall to reinitialize internal snow ``age"/state from an otherwise indeterminate value (unless $Q_0$ was specified differently in their documentation), consistent with theoretical analysis from \citeA{Wiscombe_Warren_1980}, though age/state dynamics after such initialization were dictated only by the scheme.

\subsubsection{Calibration of Existing Schemes}

Most schemes with an internal snow age ($\tau$) utilized a simplistic threshold formulation equal to the number of days since the last snowfall event ($P\Delta t > Q_0$), allowing \textit{a priori} external engineering of the $\tau$ feature and training of such schemes via regression over the same batches as the neural scheme. However, some latent-state schemes used continuous or adaptive expressions, requiring autoregressive sequence-based training. In such cases, ``batches" were constructed from contiguous cleaned data segments of at least four days. To avoid indeterminate initial latent states, new sequences were only started at days with $P\Delta t > Q_0$ liquid-equivalent snowfall. Each sequence was initialized with a new latent state and stepped forward by the parameterization using the appropriate form from Eq. (\ref{eq:diagnostic}-\ref{eq:nn_timestep}), scoring sequences by comparing simulated and observed albedos using the $L_2$ norm. All schemes could be trained on their respective batches with the same $L_2$ optimization framework. The neural scheme (and others) could also be trained in this autoregressive manner if desired, though such training was slower (roughly 5$\times$) without necessarily leading to better results compared to regression on single-day inputs and targets, leading to autoregressive training being reserved only for the models labeled ``Y" for ``AR" in Table \ref{t:models}.

\subsection{Statistical Analysis}

Evaluation metrics included root-mean-square error (RMSE), root-mean-square percentage error (RMSE\%), bias (B), and bias-percentage (B\%) between simulated and observed albedo timeseries. Statistical significance of RMSE and RMSE\% differences across validation and test sites was assessed using the Wilcoxon signed-rank test \cite{Wilcoxon_1945} at a $p = 0.05$ significance level, a nonparametric method over matched samples that avoids assumptions of normality or equal variance. The input features of the model exhibit non-negligible correlation, which can distort more traditional interpretability methods like Shapley additive explanations (SHAP) analysis, permutation importance, or local interpretable model-agnostic explanations (LIME) \cite{molnar2025}. Therefore, to qualitatively assess physical consistency, we derived a first-order accumulated local effects (ALE) plot of the neural parameterization, which isolates the marginal influence of individual features even under correlated inputs. Additional discussion of ALE construction and interpretation is available in \citeA{Charbonneau_Deck_Schneider_2025}, \citeA{Apley_Zhu_2020}, and \citeA{molnar2025}. Feature interaction strength was quantified with pairwise Friedman’s H-statistics (see Table \ref{t:interaction}), which were sufficiently small to focus analysis on primary variable behavior. Emphasis was placed on assessing larger generalization patterns and error robustness rather than isolated metric optimization or detailed mechanistic decomposition, as the study serves to introduce the potential of an adaptable parameterization framework for near-term modeling needs rather than provide an exhaustive technical analysis of a single finalized model.

\section{Results}

\subsection{Model Benchmarking}

\begin{figure}[ht!]
\noindent\includegraphics[width=\textwidth]{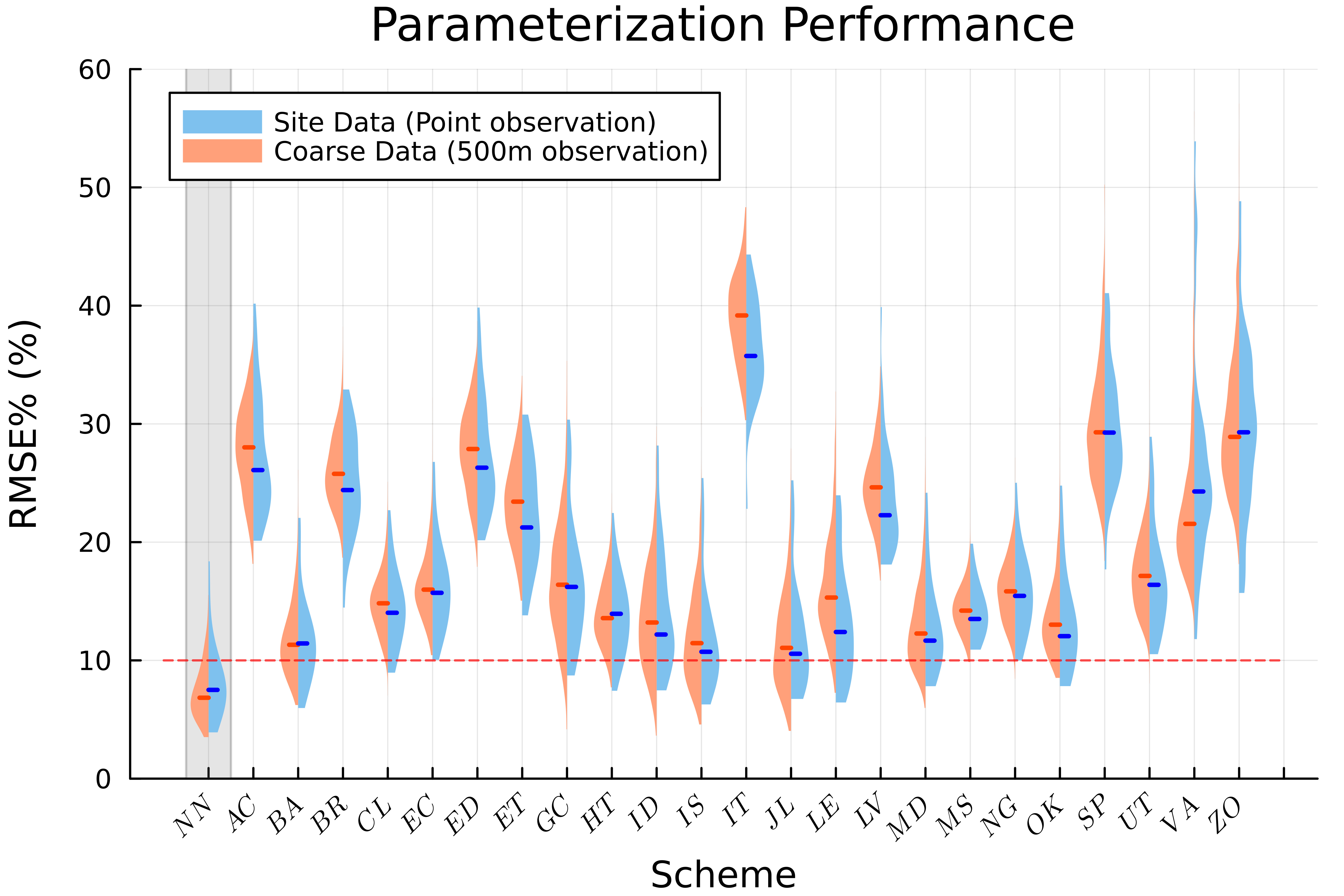}
\caption{\footnotesize \textbf{Relative Root Mean Square errors of the new parameterization ($NN$) versus existing parameterizations}. This split-violin plot shows the distribution of RMSE\% scores for each scheme across all evaluated sites (700 500m scale sites on the left in orange, 31 point observational sites on the right in blue), with the median of each set marked by the horizontal bars. Parameterization symbols match those described in Table \ref{t:models}, and our scheme is highlighted in gray to the left of the other schemes (organized alphabetically; scheme parameters taken as published). Our proposed scheme outperforms all other compared schemes, with the bulk of sites below the marked 10\% error threshold (dotted red line) common in snow albedo parameterizations. It also shows good generalizability; with a short tail (a tight spread of scores) and relatively symmetric halves between validation and testing sites (consistent performance across varied out-of-sample regimes). All scores are visible on the plot.}
\label{f:published-violins}
\end{figure}

The distribution of RMSE\% scores from all schemes as-published (various symbols, see Table \ref{t:models}) and our approach ($NN$) over both coarser (500-m scale) and site-level (point observations) data not seen during training are shown in Fig. \ref{f:published-violins}. All generated albedos by our scheme show a correlation against observations of $R^2$ = 0.786 over the coarse data and $R^2$ = 0.677 over the site-level data. The average correlation scores over each testing site can be further compared against other schemes in the modified Taylor Diagrams \cite{Elvidge_Angling_Nava_2014}, which show both standard deviations and correlations of simulations and observations (Fig. \ref{f:published-taylor}).  Table \ref{t:published-results} in Appendix B displays numerical RMSE\%, RMSE, B, B\%, and Wilcoxon Signed Rank \cite{Wilcoxon_1945} scores comparing our scheme to other schemes.

Our model exhibits a median snow albedo RMSE of 0.049 (RMSE\% of 6.8\%) across all coarse sites, increasing by less than 1\% (RMSE 0.056) for point observations, and with less than 0.01 bias at both scales, improving over 30\% from other schemes and surpassing the routine 10\% error found in other studies. The distribution of RMSE\% scores of our scheme is narrower than that of other schemes, showing a consistent range of performance, with fewer outliers (Fig. \ref{f:published-violins}). The two halves of the violin plot are relatively symmetric, suggesting good generalizability across different climates (consistency across sites) as well as generalizability across spatial scales (500 m and site-level sites). This is further corroborated by a lack of distinct patterns in performance with regards to location or snowpack type (see Fig. S1 in the Supporting Information), suggesting broad applicability. With only one calibration at coarser scales, our scheme shows an ability to readily compete against other models that are specifically calibrated for global usage across a range of locations and resolutions.

Most other schemes show less of an ability to maintain consistent performance when transferring across spatial scales --- while $NN$ shows slightly better performance on coarse data than site data, other schemes tend to show slightly better performance on site data, which could be influenced by the calibration data used for these models compared to the explicit calibration to coarse data in our approach. The lack of schemes near the unity normalized standard deviation contour in Fig. \ref{f:published-taylor} also indicates our model is more capable at replicating the variability of snow albedo observations. Our approach also shows the highest average correlation to observations. All comparisons were significant at the 5\% level ($p <$ 0.05 for all Wilcoxon tests, with 700 samples for coarse sites and 31 for point-observation sites), detailed in Table \ref{t:published-results}.

\begin{figure}[ht!]
\noindent\includegraphics[width=\textwidth]{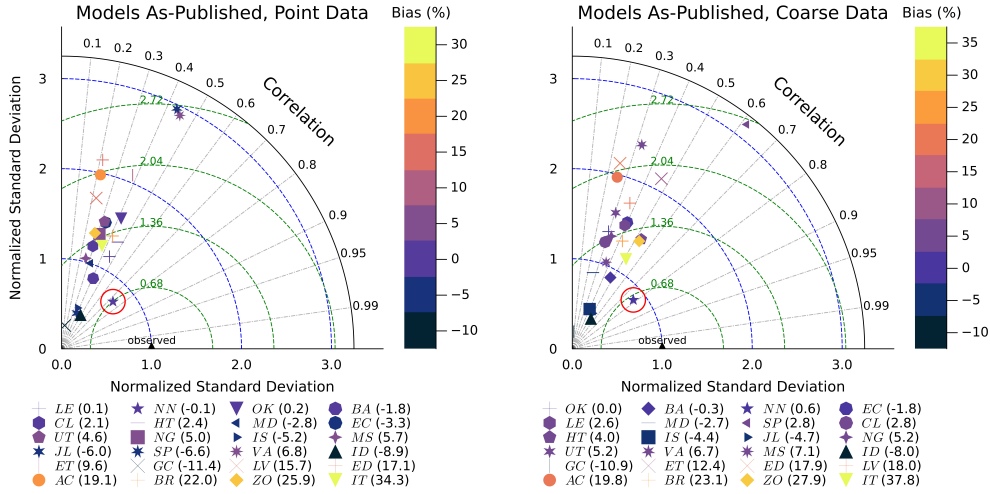}
\caption{\footnotesize \textbf{Modified Taylor Diagrams of all schemes over all ground (left) and coarse (right) testing data}. The polar coordinates of each scheme mark the standard deviation of timeseries outputs (normalized to that of the observations; the radial coordinate) and the Pearson correlation of timeseries outputs against the observations (the azimuth), averaged over all testing sites. The geometric relationship between standard deviation, correlation, and error means the Euclidean distance from each scheme’s marker to the observations is equal to the centered root-mean-squared (RMS) error against the observations (also equal to the standard deviation in model error), for which contours of centered RMS error are marked in green, contours of normalized standard deviation are marked in blue, and contours of correlation in black. Scheme markers are colored in accordance with their B\% score, and the legend is ordered in terms of increasing $|\mathrm{\%B}|$, with the value listed next to the symbol. For both types of data, our scheme (circled in red) is best able to replicate the observations with one of the best \%B scores. }
\label{f:published-taylor}
\end{figure}

\subsection{Model Consistency}

\begin{figure}[ht!]
\noindent\includegraphics[width=\textwidth]{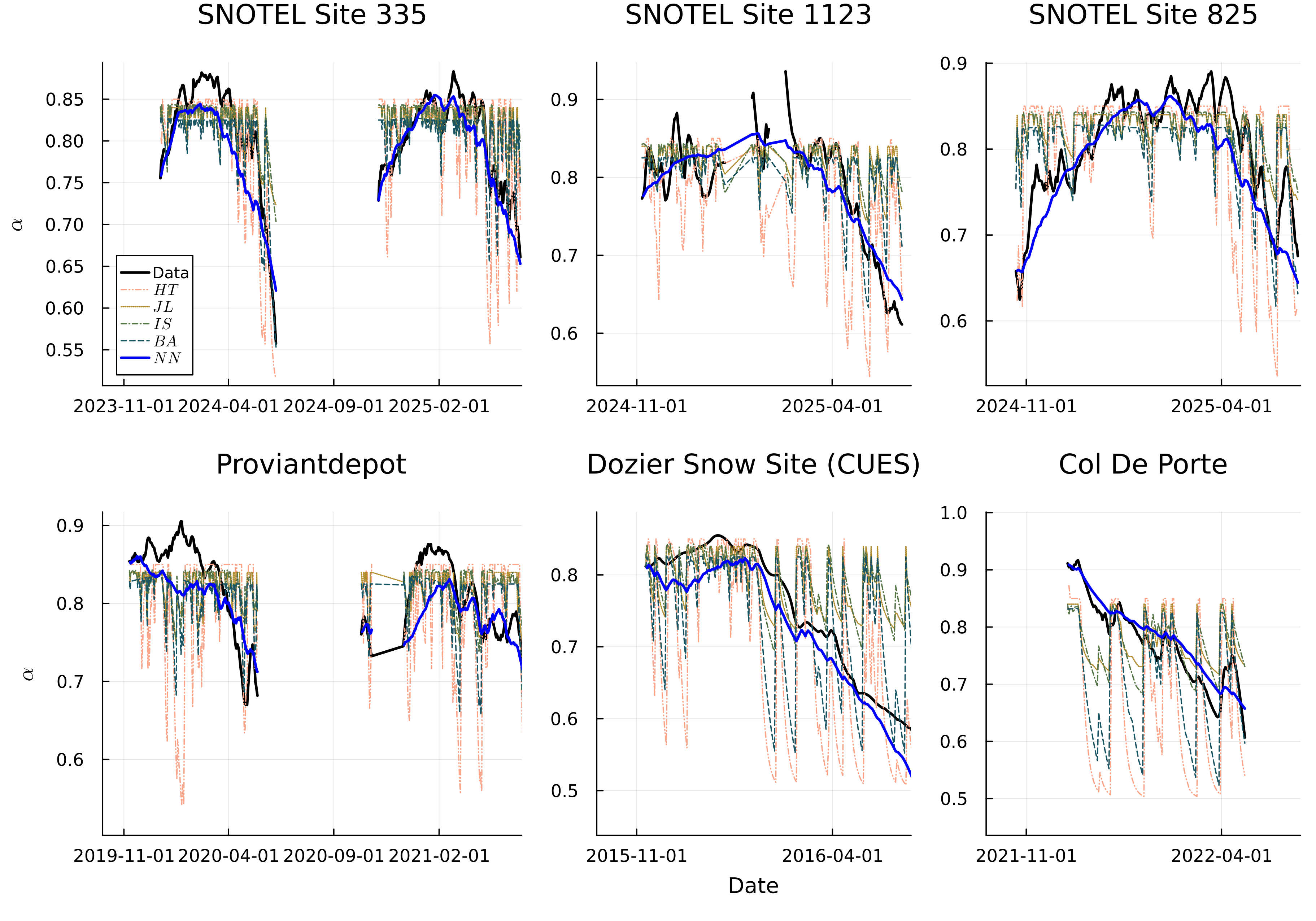}
\caption{\footnotesize \textbf{Timeseries of albedo generated by various schemes}. This figure details generated timeseries over a selection of site-level testing sites compared to the processed timeseries data, using symbols provided in Table \ref{t:models}. While other schemes show a better capability at creating fast fluctuations or jumps from snowfall, our scheme shows a better ability to follow the overall trend of the albedo throughout the snow season. The shown years were selected to highlight the difference between the trend-tracking behavior of our model vs the fluctuations of other parameterizations.}
\label{f:timeseries}
\end{figure}

Timeseries of our scheme output versus that of well-known top-quintile alternative schemes like BATS (BA), JULES (JL), ISBA (IS), and HTESSEL (HT) are given in Fig. \ref{f:timeseries}. Compared to the other schemes, our approach is more able to trace the trends of the albedo within the growth and melt season and match the range of slopes created by the observations across different conditions. By expressing a variety of growth rates, the neural model can better track the seasonal dynamics compared with other models with more rigid evolution structures or a finite set of growth and depletion rates. While this is useful over accumulated seasonal timescales (and thus useful for gauging the accuracy of accumulated energy fluxes within a GCM), we do note this approach does not capture the abrupt increase in albedo due to new snowfall or fast decreases as well as the other schemes, preferring to instead more smoothly climb and fall and transition between different growth and depletion rates, which is likely influenced by our use of training data at coarser scales, $L_2$ loss, and data cleaning approaches. At higher temporal resolutions where high-frequency albedo changes can occur from snowfall or intraday zenith angle dependencies, our approach will likely track fluctuations less but track the average well. This emphasis is more conducive to the needs of climate models and long-term studies, where low systematic bias in trends and integrated fluxes is critical over sub-weekly variation or transient fluctuations \cite{CLIM, NGUYEN2016117}.

\begin{figure}[ht!]
\noindent\includegraphics[width=\textwidth]{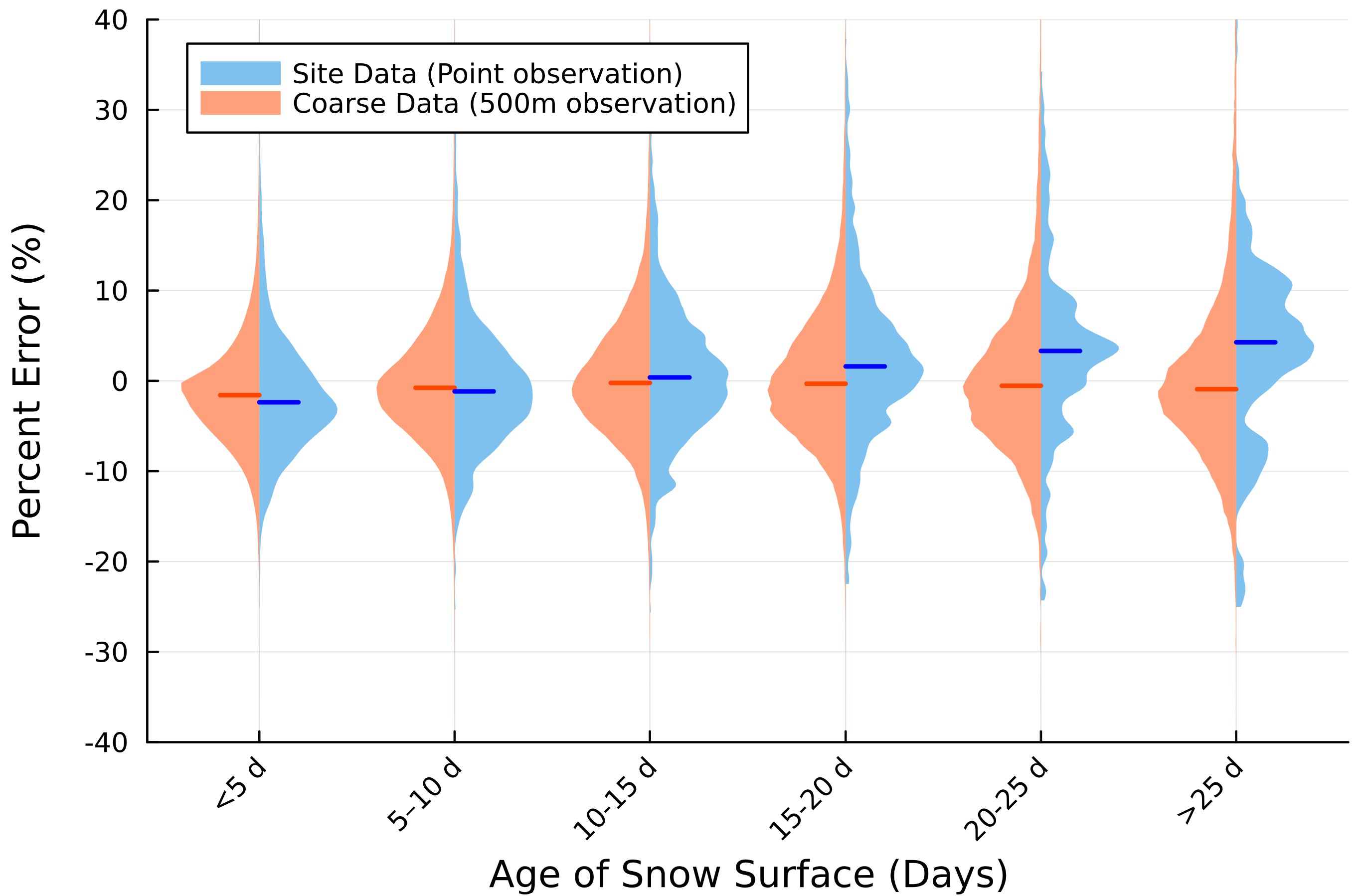}
\caption{\footnotesize \textbf{Timeseries error decomposition by snow surface age}. This figure details the breakdown of RMSE\% scores in generated timeseries by the age of the snow surface for the developed ML scheme, following the split-violin format introduced in Fig. \ref{f:published-violins}. The symmetry between coarse and site-level results breaks when snow age is more than 2 weeks old, with higher error for older snow surfaces, though all errors still remain small compared to those found in the other schemes.}
\label{f:age_breakdown}
\end{figure}

If the spread of scores aggregated across snow seasons was primarily dominated only by strong seasonal signals (i.e., merely reproducing the average increase-decrease cycle from the training distribution), strong degradation and patterns should be expected when transferring the scheme between regimes of different climatic and snow conditions. To investigate this, timeseries errors decomposed by snow surface age are displayed in Fig. \ref{f:age_breakdown}, which additionally allows exploration of snow errors within a snow season and by snow type. A similar break-down of error spread by snow year, snow season average temperature, and season total snow accumulation were also assessed (see Fig. S2--S4 in the Supporting Information), exploring the scheme robustness against more extreme or out-of-distribution conditions.

For error decomposition by snow age, the difference between coarse and site-level performance is more separated for older snow surfaces. Error increases slightly with snow age for site-level evaluations, but the majority of errors still remain under the 10\% level, meaning the spread in errors even with snow age effects remains small. For coarse-level data, the distribution of errors remains the same regardless of snow age, and errors between coarse and site-level data remain symmetric until the snow surface is roughly 2 weeks old. These similarities again suggest the developed simple scheme exhibits capable generalization between snow types and across spatial scales. Results of the other error decompositions by snow year, average temperature, or total snowfall exhibited even weaker patterns (Fig S2--S4 in the Supporting Information), with little pattern or variance in error spread across a variety of snow age or environmental conditions, particularly for the coarse-level data. Although seasonal evolution contributes substantially to snow albedo variability, the model is trained only on daily albedo tendencies independent of the bigger snow season, and exhibits relatively uniform performance across geographically distinct sites (Fig. S1) in addition to diverse snow and environmental regimes. This consistency implies that model skill is not solely derived from reproducing a generic seasonal cycle and exhibits capable generalization across a wide variety of evaluation modes from a single training.

To assess the scheme's behavior by-feature, an Accumulated Local Effects (ALE) plot of our scheme is shown in Fig. \ref{f:ale} below, with average outputs and errors under different environmental conditions tabulated in Table \ref{t:outputs}. The ALE plot displays the centered changes in model output (the change in output $d\alpha / dt$ compared to the average output $d\alpha / dt$), accumulated over sequential bins of feature values. This mitigates some of the distortions that can arise in marginal effect analyses when predictors are correlated, making it more suitable than partial dependence methods for interpreting learned responses in this setting where predictors are not fully independent. Interpretation of curve shape (but less so scale) can then be used to gauge behavioral alignment with physically-informed expectations. In addition to the curve for the presented scheme, the distribution of curves from another 50 trained schemes (identical structure, different initial weights) is also displayed to visualize scheme sensitivity. 

\begin{figure}[ht!]
\noindent\includegraphics[width=\textwidth]{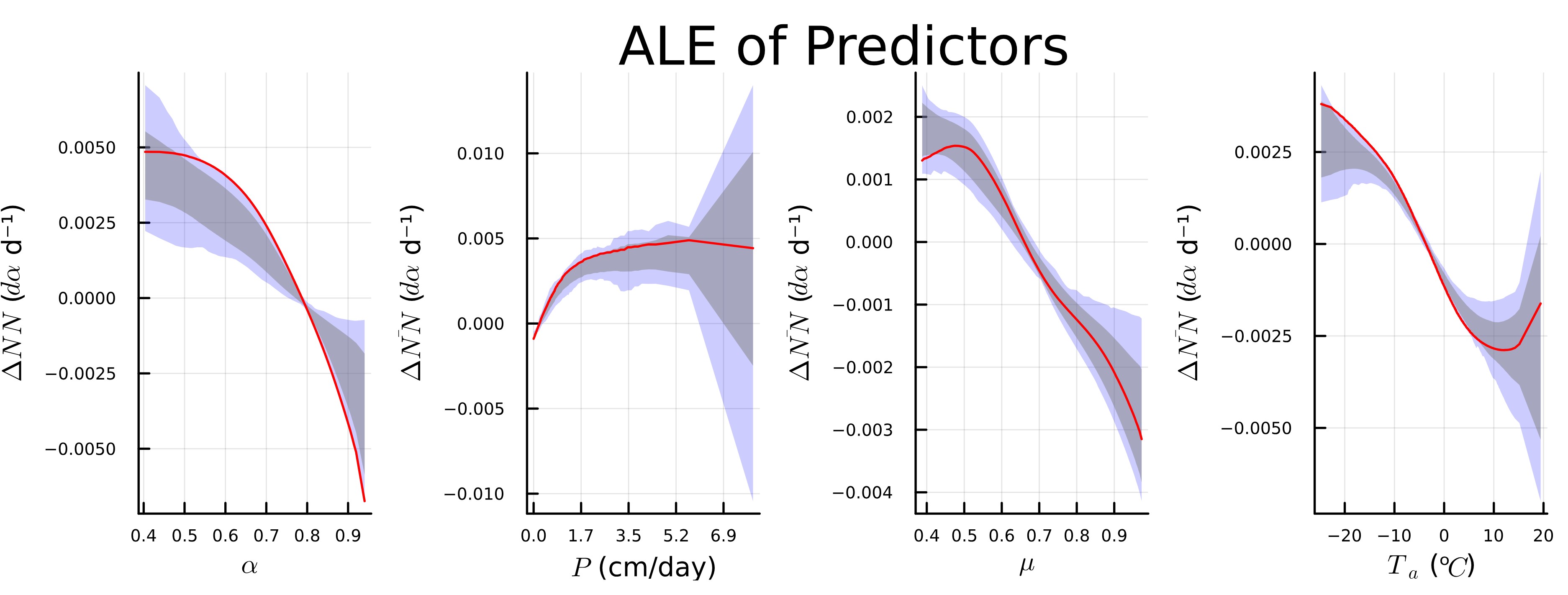}
\caption{\footnotesize \textbf{ALE Plot for features of $NN$}. The horizontal axes give the range of each of the four input variables (present albedo $\alpha$, water-equivalent snowfall $P$, cosine of solar noon zenith angle $\mu$, and air  temperature $T_a$) while the vertical axes show the average change in the scheme relative to the average prediction of albedo change per day (which is $d\alpha /dt \approx 0$). Curves are binned so that each bin has at least 50 samples to average. Gray bands represent the mean and one standard deviation of the ALEs from an ensemble of trained models, while the blue bands fill the 5\%-95\% quantiles. Scheme behavior becomes more variable in regions where there are limited samples to inform the training process, towards the extremes of the available data (e.g., $P$ after 5.5 cm d$^{-1}$).}
\label{f:ale}
\end{figure}

\begin{table*}[ht!]
\caption{\footnotesize \textbf{Quantified Scheme Behavior}. This table displays outputs and errors from the scheme over both the coarse and site level testing data across a range of environmental conditions.}
\label{t:outputs}
\begin{center}
\begin{tabular}{c|cc}
\hline
Values & Coarse & Site \\ \hline
Mean $\mathrm{|RMSE\%|}$ & 4.98\% & 5.79\% \\
Mean $\mathrm{|RMSE\%|}$, Snowing & 4.88\% & 5.57\% \\
Mean $\mathrm{|RMSE\%|}$, No Snow & 5.05\% & 5.85\% \\
Mean $\mathrm{|RMSE\%|}$, No Snow, $T_a > 0^\circ$C & 5.86\% & 6.38\% \\
Mean $\mathrm{|RMSE\%|}$, No Snow, $T_a < 0^\circ$C & 4.55\% & 5.49\% \\
Mean $d\alpha / dt$, Snowing & 0.00085 $\mathrm{d^{-1}}$ & 0.0012 $\mathrm{d^{-1}}$  \\
Mean $d\alpha / dt$, No Snow & -0.0021 $\mathrm{d^{-1}}$  & -0.0016 $\mathrm{d^{-1}}$  \\
Mean $d\alpha / dt$, No Snow, $T_a > 0^\circ$C & -0.0036 $\mathrm{d^{-1}}$  & -0.0031 $\mathrm{d^{-1}}$  \\
Mean $d\alpha / dt$, No Snow, $T_a < 0^\circ$C & -0.0013 $\mathrm{d^{-1}}$  & -0.00067 $\mathrm{d^{-1}}$  \\ \hline
\end{tabular}%
\end{center}
\end{table*}

The RMSE\% errors in Table \ref{t:outputs} for both coarse and site-level data vary by only $\sim$1\% across a range of environmental conditions, without accumulation in a seasonal manner. Despite the model failing to capture abrupt increases from snowfall compared to other schemes and a low average output derivative during snowfall events ($\sim$0.001 per day), the average error between snowfall and no-snowfall events remains similar, though it is slightly higher for days without snow that are above the melting temperature than those below the melting temperature. The output derivatives by the network are also positive under snow conditions and negative under no-snow conditions, with larger decreases for above-freezing $T_a$ than below-freezing $T_a$, aligning with expected physical behavior. The output derivatives without precipitation average roughly -0.002 $\mathrm{d^{-1}}$, which is similar order to (and about half the magnitude, though the means are averaged over a range of values) those found in analogous linear albedo decay equations like that of \citeA{Baker_Ruschy_Wall_1990} or the schemes in UTOPIA (UT) and HTESSEL (HT). The slightly higher error under warmer temperatures and no precipitation corroborates the findings with snow age in Fig. \ref{f:age_breakdown}, suggesting the model performs slightly better in early season than in late season like many established albedo models.

The mean prediction of our model over all data is $d\alpha / dt \approx 0$, so positive/negative y-axis values in each ALE plot (the average change in model output relative to the mean prediction) can be interpreted as the model outputting positive/negative derivatives, as opposed to merely more positive/negative values than some finite average. The ALE curves suggest that the learned statistical relationships are broadly consistent with established physical expectations for snow albedo evolution (see Fig. \ref{f:ale}). Albedo tends to decrease unless there is a minimal amount of snowfall (2.1 mm liquid-equivalent), which then increases but levels off as the amount of snowfall increases past a $\sim2$ cm/day threshold. The minimal snowfall amount is consistent with the utilized data resolution (2.54 mm), and the  stabilization threshold is consistent with measured limits of optically thick snowpacks for both old and new grain sizes \cite{dozier1991retrieval, Wiscombe_Warren_1980}. Similarly, the model begins to tend towards predicting decreases in albedo as temperatures approach and increase past $-3.6^\circ$C, with a slower rate of change in output near colder temperatures where less melting would occur. This value is sufficiently close to the melting temperature where we would expect melt-based albedo decay to manifest, and is consistent with the temperature threshold found in HTESSEL ($-2^\circ$C) to dictate albedo decay. The albedo curve exhibits regression to the mean, with an intercept of 0.78, the mean albedo in the training data.

While intraday albedo changes decrease and then increase (an intercept of $\mu = 0.66$, about 48 degrees, which is also the mean in the training data) as the sun enters and exits at low zenith angles, the daily albedo trained here uses the zenith angle cosine at local noon, which increases during warmer seasons. The ALE plot thus shows the tendency for albedo to decrease during warmer days of the year, again consistent with physical expectations. However, observational or structural limitations can influence these results --- the concavity at higher $T_a$ could indicate an artifact of the training, the slowing rate of albedo decrease as albedo approaches its minimal values due to low/no snowpack at the temperature ranges shown, or could alternatively be influenced by the number of samples in the bins at the extreme ranges of the data (there are orders of magnitude less samples after $T_a >$ 15 $^\circ $C or after P $>$ 6 cm/day compared to elsewhere along the curve, so statistics can become slightly skewed or the curve become less smooth compared to other feature ranges). The estimation uncertainty of each ALE curve in the ensemble (the standard errors of the means that build the curve, accumulated in quadrature when the local effects are accumulated) is negligible (less than the width of the line on the plot) compared to the range of curves across different trained schemes, implying the structural/parameter uncertainty of the presented model form is dominant. This means the spread seen in Fig. \ref{f:ale} is indicative of a high sensitivity to the scheme initial state, though this is unsurprising for a small neural network. The stability of the learned relationships, however, is comparatively robust within most of the ranges found in the data, with the ALE ranges varying more at extreme values (which are poorly represented in the data and thus carry comparatively little influence on the training process) but the overall shapes and scales of the curves remain similar. We note that the reported interpretability results should be viewed as diagnostic rather than mechanistic or implicative of causality. The ALE analysis characterizes the statistical sensitivities learned by the network and does not establish causal physical relationships, though it permits exploration of their alignment with expected behavior. 

\subsection{Calibrated Results}

\begin{figure}[ht!]
\noindent\includegraphics[width=\textwidth]{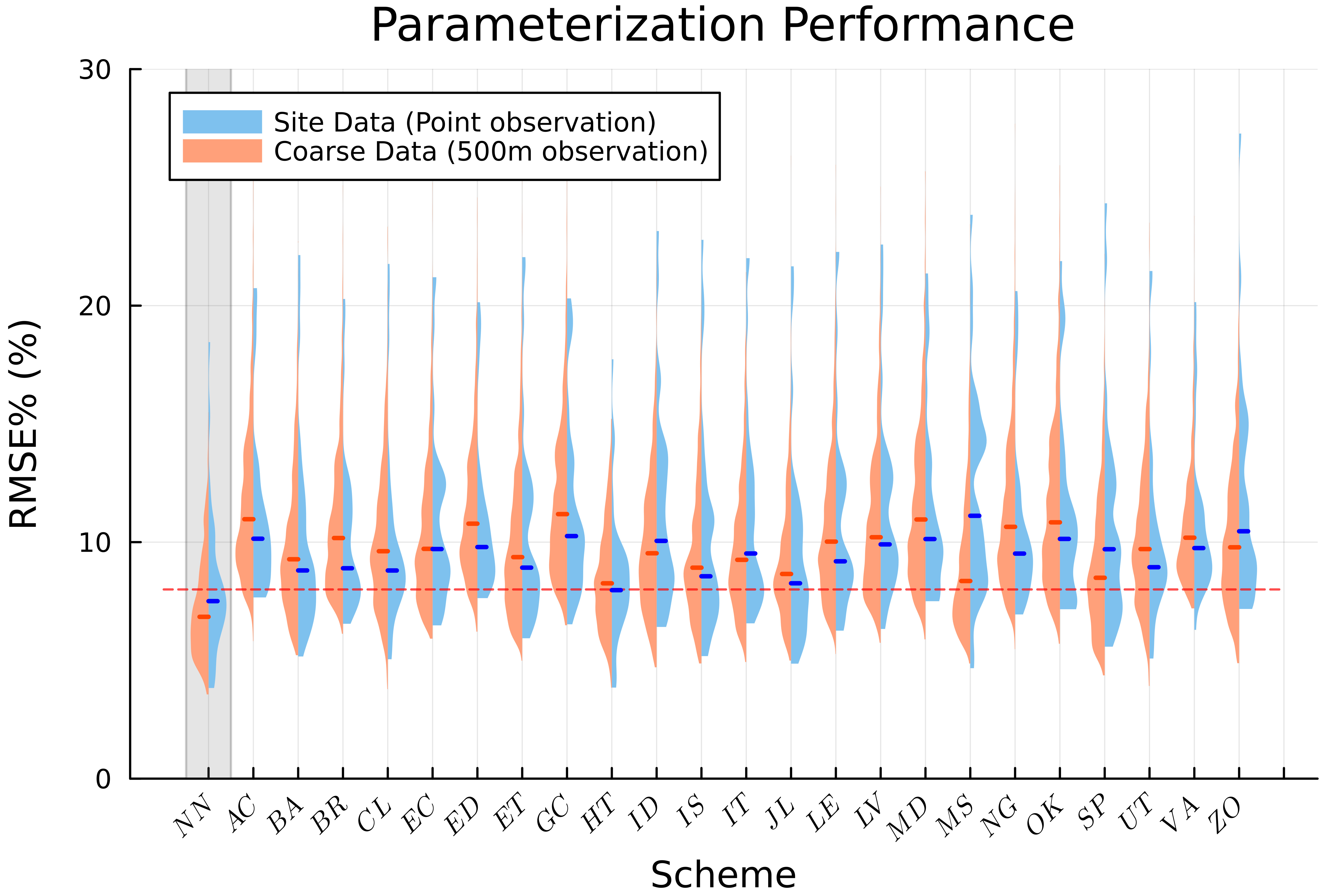}
\caption{\footnotesize \textbf{RMSE\% scores of all calibrated parameterizations}. Format follows that seen in Fig. \ref{f:published-violins}, though this time the red dotted line is placed at an RMSE\% of 8\% instead of 10\% to help better distinguish scheme performance. Our scheme remains clearly separated from other schemes.}
\label{f:calibrated-violins}
\end{figure}

\begin{figure}[ht!]
\noindent\includegraphics[width=\textwidth]{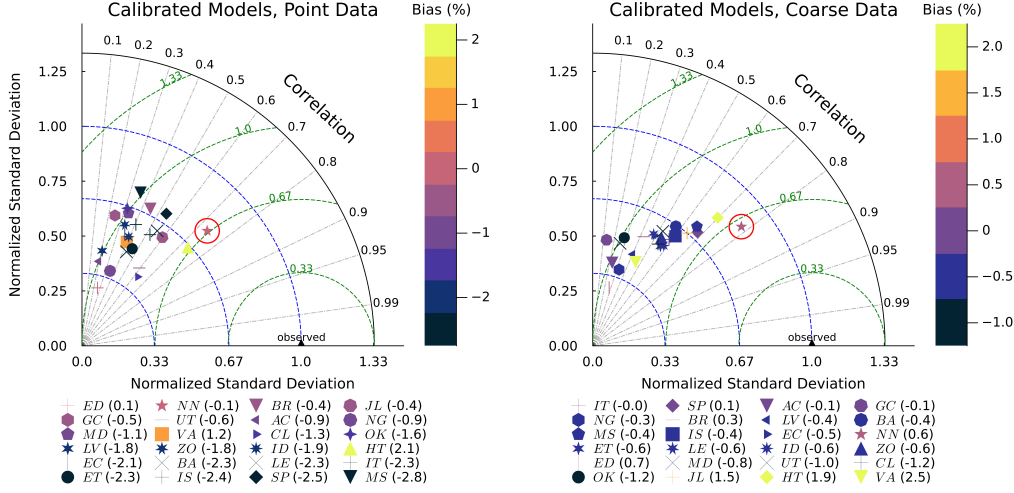}
\caption{\footnotesize \textbf{Modified Taylor Diagrams of Calibrated parameterizations}. Format follows that seen in Fig. \ref{f:published-taylor}. Our model still exhibits the highest average correlation and is the closest to the observed values, though the spread is unsurprisingly tighter than in the uncalibrated case. Our scheme also shows a stronger ability than the other models to match the observational variation in albedo, and is still among the better performers with regards to bias, especially at point scales.}
\label{f:calibrated-taylor}
\end{figure}

A ``best-case" ancillary consideration for holistic comparison against our scheme was also examined, by additionally calibrating all other schemes (already designed for global usage) directly to the data to examine the best performance their functional form could achieve (see Overview in the Supporting Information). Calibrating specifically to the same data did not change the overall conclusions and rough scheme ranking, in alignment with findings from \citeA{Magnusson_Wever_Essery_Helbig_Winstral_Jonas_2015} and \citeA{Essery_Morin_Lejeune_B_Menard_2013}. Many other schemes still show a large number of sites with errors in the 10-20\% range after calibration (see Fig. \ref{f:calibrated-violins}), while the bulk of sites at both scales by our model are in the $<$ 10\% error range. Even after calibrating the other schemes (maintaining our scheme’s original calibration), our model exhibited significant improvement in RMSE and RMSE\% at the 5\% level for both coarse and point resolution, except for one insignificant RMSE\% score ($p \approx$ 0.21) yet a near-significant improvement in the corresponding RMSE score ($p = 0.0695$) against the HTESSEL scheme (HT), for only the point-level data – our model still showed significant improvement over HTESSEL at the coarse level. Despite this, our scheme still showed roughly 0.5 RMSE\% (RMSE 0.007) better performance over point-level sites than HTESSEL, and reduced bias by nearly an order of magnitude (see Table \ref{t:calibrated-results} in Appendix B). The BATS (BA) and JULES (JL) schemes were still among the better schemes after calibration, though the HTESSEL (HT) scheme increased in competitive ranking. That our model and other prognostic schemes are among the top performers corroborates earlier findings from \citeA{Essery_Morin_Lejeune_B_Menard_2013}, implying prognostic schemes fare better than diagnostic schemes. Even after calibration to the same data, most other schemes still exhibit better performance at the site level than the coarse level - this could be due to their tendency to fluctuate more than the $NN$, impacting their ability to emphasize the lower-frequency trends that dominate at coarse-grained scales. Fig. \ref{f:calibrated-taylor} shows that calibration aided all schemes in reproducing the variability seen in the data, though our proposed scheme continues to be the best at reproducing the observations, having the second lowest mean B\% and top quintile of median B\% for point observations. All scheme equations and both their published and calibrated parameters are provided in the Supporting Information.

\section{Discussion}
This study provides a proof-of-concept demonstration and initial benchmark of a constrained machine-learning parameterization framework for snow albedo evolution, which focuses on deterministic prediction skill and generalization. With only one-time training and a simple model structure, our parameterization improves simulation accuracy by 10–30\% relative to standard schemes. The approach demonstrates transferability across a broad set of contemporary climates, spatial scales, and snow/environmental conditions, at levels unattained by existing analogous schemes. While the different evaluation data do not cover all possible snow environments nor provide inputs to explicitly benchmark across varying snow types, we observe broad generalization across a spread of geographically and climatically distinct regimes and snow states. This performance consistency indicates the approach is extracting relationships broadly consistent with physical expectations from the input environmental and snow state data that can translate across scales and environmental regimes. The results mirror our earlier work developing a similar neural model for snow thickness \cite{Charbonneau_Deck_Schneider_2025}, reinforcing the promise of the utilized framework for usage in ever-finer-resolution GCMs. 

Despite its simplicity, the network exhibits accuracy and transferability which is beneficial for consistency at global scales and fidelity in local applications. The established generalization by this scheme suggests a robustness against nonstationarity and long-term forecasts as climate regimes shift over time, compared to other parameterizations more sensitive to their calibration regime, though explicit evaluation of extrapolation under future or strongly out-of-distribution conditions remains an important direction for future work. The minimal variable set also decouples the output albedo from compounding or systematic errors in other simulated or uncertain inputs that are necessary in alternative schemes, further aiding robustness. Performance levels are within the 0.05-0.08 RMSE ranges of more complex schemes from radiative-transfer theory like SNICAR \cite{Shao_Xu_Li_Wang_Hao_2020} or full physical models such as Crocus (using the IS scheme) \cite{Gaillard_Vionnet_Lafaysse_Dumont_Ginoux_2025, Alexander_Tedesco_Fettweis_van_de_Wal_Smeets_van_den_Broeke_2014} when optimized within a simulation. In real simulation contexts, where coarser scales and uncertain inputs degrade accuracy, these advanced models typically exceed 0.1 RMSE \cite{Hao_Bisht_Rittger_Stillinger_Bair_Gu_Leung_2023, Lin_He_Abolafia-Rosenzweig_Chen_Wang_Barlage_Gochis_2025, Oaida_Xue_Flanner_Skiles_De_Sales_Painter_2015, Painter_Bryant_Skiles_2012}, whereas our scheme will not require such intermediate modeling inputs. 

The model also demonstrates competitive skill within the ML space. To our knowledge, this is the first application of ML specifically for the parameterization of snow albedo from meteorological inputs, which can be applied within a GCM for future albedo simulation. Previous work like that of \citeA{Ding_Liang_Ma_He_Jia_Wang_2024} retrieved snow albedo using non-state inputs such as day of year, top-of-atmosphere reflectance values, and satellite band ratios, but also reanalysis-derived albedo, or that of \citeA{Zhao_Ding_Wang_Qu_2025} which used gradient-boosted trees and other ML models to predict snow surface albedo from satellite surface reflectance values, trained using simulated data from an atmospheric radiative transfer model. Both studies found their designs improved prediction consistency against site-level observations, achieving mean RMSE $\sim 0.055-0.075$ (about 7.5-10.3\%), over 101 worldwide sites or 26 sites in Greenland. However, by relying on reanalysis or satellite inputs, it cannot forecast albedo forward in time from the present within coupled systems, limiting the approach to post-processing instead of prognostic use. Comparable ML efforts (e.g., \citeNP{Chen_Xiao_Zhang_Pellikka_Liu_Liu_2025, Kiem_Hammerle_Montagnani_Wohlfahrt_2024, Jia_Zeng_Tian_Chen_Chen_Wang_Wang_Ding_2025}) and statistical approaches (e.g., \citeNP{Ye_Cheng_Hao_Yu_Ma_Liang_Shen_2023, Jia_Wang_Liang_Peng_Yu_2023}) achieve similar accuracy but rely on geographic indices or predictors such as satellite band ratios, which constrain resolution and applicability and reduce physical interpretability, Our smaller, simpler scheme achieves similar results to these state-of-the-art studies and other ML techniques, while trained on only $\sim$5\% of available samples and only meterological inputs, improving interpretability and long-term applicability. \citeA{chevro} trained a neural network to emulate SNICAR along with a network inversion technique to additionally infer surface properties from input albedo spectra. Their approach achieved substantial speed-up relative to the full radiative transfer model while diagnosing spectrally resolved albedo (212 bands) with albedo errors below 0.01. However, the emulator needed over 5 million training samples to train 12 layers and 266k parameters, orders of magnitude more parameters than entire GCMs; limiting feasibility for online fine-tuning within GCM ensembles. As with the SNICAR model they emulate, their performance depends on precise input availability (snow grain size, liquid water content, pollutant concentrations, or full 212-band spectra for inversion), requirements unlikely to be met in near-term climate modeling. The reported cost of inversion to obtain such surface properties ($\sim$0.1 s per grid point) likewise severely constrains scalability for global models. In contrast, our standalone albedo dynamics model is nearly 1000x smaller, enabling online tuning and rapid prototyping, and achieves our results with 100x less data, requiring under 2 $\mu$s per grid point. More notably, it does not inherit the assumptions in an underlying model (e.g., constant snow density and semi-infinite snow depth for the SNICAR emulator), thereby avoiding associated downstream biases.  

The model outcomes are constrained by the context, observational uncertainty, and limitations of the available data, as any data-driven model will inherit characteristics of its training data. The SNOTEL network and site-level observations capture localized relationships but are noisy, prone to quality issues, and do not represent large-scale heterogeneity, particularly in complex terrain \cite{Meyer_Jin_Wang_2012, Hill_Burakowski_Crumley_Keon_Hu_Arendt_Wikstrom_Jones_Wolken_2019}. The STC/MODSCAG-MODDRFS product likewise suffers terrain as well as cloud effects and is less accurate in forests, partly because its reference spectra exclude burned vegetation. This omission is noteworthy as soot- or fire-derived LAPs increasingly drive abrupt albedo darkening in forested regions, where most SNOTEL sensors reside \cite{Jensen_Rittger_Raleigh_2024}. Both data sources are affected by persistent obfuscation from debris, vegetation, or shadows, and lack coverage in extreme snow environments like tundra and taiga biomes, leaving our model’s extension to these domains untested. The utilized SNOTEL sites express a range of possible LAP environments, but are not wholly encompassing, so it is possible that areas with more extreme sand or soot content could yield biased model estimates. Our simulation errors are already reaching the observational uncertainty range of modern data efforts themselves, underscoring the effectiveness of the modeling paradigm but implying further model gains are unlikely without significant advances in observational quality and availability. Future retraining with improved products is straightforward within the proposed framework.

The tradeoff between smoothing noisy timeseries and preserving short-lived fluctuations is difficult to avoid. We favor smoothing in our methodology (see Appendix A for more detail), which provides a stable representation of the trends relevant for climate modeling, though this can blur the true input-output relationships; potentially limiting model extension to sub-daily scale variability without retraining. Separability of learnable albedo responses is also restricted by the chosen inputs, which cannot capture LAP concentrations, wildfires \cite{Hatchett_Koshkin_Guirguis_Rittger_Nolin_Heggli_Rhoades_East_Siirila_Woodburn_Brandt_et_al._2023}, grain size, snow chemistry \cite{Sarangi_Qian_Rittger_Bormann_Liu_Wang_Wan_Lin_Painter_2019}, or atmospheric state. Aggregating these influences into one broadband albedo, rather than spectral or direct/diffuse components, compounds this challenge, especially near more extreme $d\alpha / dt$. While additional variables could represent more processes, direct measurement of such effects is rare, uncertain, or infeasible at GCM scales \cite{Gunther_Marke_Essery_Strasser_2019, Bair_Davis_Dozier_2018, Painter_Berisford_Boardman_Bormann_Deems_Gehrke_Hedrick_Joyce_Laidlaw_Marks_et_al._2016}. Our framework, however, can readily adapt if such data becomes available. 

This work was designed to evaluate predictive performance and transferability across scales for the proposed framework, but does not aim to provide one operational design or formal uncertainty quantification of a finalized model. Several sources of uncertainty are therefore intertwined in the reported results. In addition to the aforementioned observational uncertainty arising from measurement error, quality-control procedure, and retrieval availability, representativeness differences between point and gridded observations can impact the derived results. Structural uncertainty arises from the parameterization itself, including feature selection and omission, network architecture, and temporal resolution. Additional uncertainty may arise from the imposed bounds, which were selected to improve stability and realism while minimizing assumptions, but may contribute to representation errors under extreme conditions and do not inherently prohibit inconsistent behavior between the limits. The distributions of model errors across sites provide an indirect indication of performance variability, but should not be interpreted as formal prediction intervals. Future work could include ensemble, Bayesian, or probabilistic formulations to estimate uncertainty and propagate observational uncertainty through the parameterization, as the framework remains adaptable to such techniques.

The model captures a diverse range of $d\alpha / dt$ well but struggles with sharp increases from new snowfall at the site level. This limitation likely reflects data separability issues for extreme $d\alpha / dt$ from both the input variable choice (the feature values associated with extreme $d\alpha /dt$ are also associated with other non-extreme targets within the spread of data) and the cleaning techniques, as well as the selected $L_2$ loss, which emphasizes long-term statistics over individual predictions. Prognostic schemes also require an initial albedo value at the onset of a new snowpack, which can vary by environment \cite{Abolafia_Rosenzweig_He_McKenzie_Skiles_Chen_Gochis_2022}, leading to spin-up effects. However, high-frequency albedo variation at point-scales will be suppressed in the coarse-graining over grid cells within a GCM. While replicating extreme and sub-weekly variability are important for event-level and point-scale studies in the earth sciences, seasonal energy balance and time-integrated radiative effects dominate evolution within an earth system model and long-term simulations, as these govern large-scale feedbacks and equilibrium responses \cite{Qu_Hall, CLIM}. This leaves our computationally efficient approach well-suited for GCM integration. A small network structure can approximate a variety of nonlinear functions while maintaining linear computational scalability. This also lowers overfitting risk while evading the rigidity of symbolic functions with even fewer parameters. This contrasts with competing age-based parameterizations, which rely on precise knowledge of snowfall timing that are uncertain at large scales \cite{Bair_Rittger_Skiles_Dozier_2019}. The proposed scheme adapts better to inherent observational uncertainty, enables rapid iteration, and can readily incorporate new inputs, motivating broader adoption as observational capacity expands. 

We note that behavior consistent with physical expectations of precipitation effects was learned even without choosing boundary functions to explicitly enforce it, illustrating the potential for this approach. However, the absence of explicit physical constraints may increase predictive uncertainty when extrapolating beyond the environmental conditions found in the training and testing data. While the developed scheme demonstrates strong transferability across locations and scales, explicit evaluation under extreme out-of-distribution or future climate states and among features that can quantify radiative transfer effects remains an important avenue for future work. Further experimentation with alternative analytical boundaries could also enforce other desirable behaviors or further increase consistency beyond this initial investigation. 

The parameterization performs well at point scales and daily timeseries from 500 m resolution. Still, these spatial resolutions are finer than the resolutions of current GCMs, where even more coarse-graining or shorter time-steps may introduce additional variability. However, the continuing push for finer-resolution highlights our approach as a candidate for the growing needs of next-generation climate models. Benchmarking accuracy and spatial structure within a full model over contemporaneous spatial grids at global scale or by land type was not assessed here, and is being addressed in an upcoming paper where the prognostic model has been integrated into a new land surface model \cite{climaland} at sub-daily resolution. Other additional avenues for research include other temporal dependencies, validation across global flux tower networks, or adapting the framework for spectrally resolved albedo using alternative MODIS products (e.g., MCD43D) to compare against more complex parameterizations.

\section{Conclusion}

By developing a simple prognostic parameterization for snow-albedo change that breaks from a continued reliance on snow-age, we simulated seasonal daily snow albedo with a median relative error of under 7.5\% across diverse locations (median RMSE of $\sim 0.05$ at both coarse and point scales). Training over a wide data set creates more generalizable results. Our performance extends from point-scale to coarse-grained data and multiple climate types; outperforming or matching advanced data products, complex physical models, and contemporary ML approaches using only standard meterological inputs at low computational burden. This reinforces the potential of a constrained data-driven framework that preserves physical consistency while offering accuracy, scalability, and adaptability for next-generation Earth system models. The approach can be used within larger models and can accommodate evolving observational availability. Because snow cover at different altitudes and locations samples situations that may arise in a future, warmer climate (where snowlines generally move to higher altitudes and latitudes), the generalizability shown in this study suggests potential robustness under changing climate conditions, which can be quantified more rigorously in future study. This application of constrained ML for snow albedo modeling shows potential for developing more accurate forecasts and effective mitigation strategies in an evolving climate, and offers a flexible path for integrating ML into physical modeling without compromising interpretability or physical consistency.

\appendix
\section{Data Appendix}
\subsection{Remote Data}
Many remotely-sensed snow albedo products exist, with MODIS-derived ($\sim$500 m, daily) products remaining standard despite the emergence of finer but less frequent ($<$250 m) imagery. The STC/MODSCAG-MODDRFS 500-m albedo product (hereafter ``SMM") product \cite{Rittger_Raleigh_Dozier_Hill_Lutz_Painter_2020, Painter_Rittger_McKenzie_Slaughter_Davis_Dozier_2009, Painter_Bryant_Skiles_2012, RITTGER_Lenard_Palomaki_Brodzik_Stillinger_Bair_Dozier_Painter_2024, Jensen_Rittger_Raleigh_2024} applies spectral unmixing of MODIS reflectances against simulated radiative-transfer and observational libraries to isolate snow albedo, outperforming empirical products like MOD10A1 that more readily represent the mixed total land surface albedo \cite{Fleming_Rittger_Oaida_Taglialatela_Graczyk_2024}. SMM has been recommended for next-generation model calibration \cite{Fleming_Rittger_Oaida_Taglialatela_Graczyk_2024}, though many studies still rely on outdated products, impairing parameterization accuracy and performance interpretability \cite{Aguirre_Bozkurt_Sauter_Carrasco_Schneider_Jana_Casassa_2023}. SMM ownership recently transferred to Snow Today via the National Snow and Ice Data Center (NSIDC). 

Work continues on improving known limitations in SMM data, including pixel geolocation errors ($\sim$1 pixel), high solar-zenith-angle albedos, vegetation/cloud/topographic interference, spectral library representativeness, and inherent one-to-many ambiguity in the unmixing algorithm \cite{Hao_Bisht_Rittger_Stillinger_Bair_Gu_Leung_2023, Rittger_Raleigh_Dozier_Hill_Lutz_Painter_2020, Bair_Rittger_Skiles_Dozier_2019, Jensen_Rittger_Raleigh_2024, Demil_Haghighi_Klove_Oussalah_2025, Keuris_Hetzenecker_Nagler_Molg_Schwaizer_2023}. Sensor drift in the Terra/Aqua MODIS sensors also introduces minor bias \cite{Feng_Wehrle_Cook_Anesio_Box_Benning_Tranter_2024}, though effects are minimal among the temporal ranges and quality measures present in SMM and this study. NSIDC recently prioritized SPIReS, a newer unmixing algorithm \cite{rittger2023coproduction}, but we used SMM due to broader validation, minor differences in accuracy \cite{Hao_Bisht_Rittger_Stillinger_Bair_Gu_Leung_2023, Stillinger_Rittger_Raleigh_Michell_Davis_Bair_2023}, and better availability during much of this study.

\subsection{In-Situ Data}

In-situ data measure collocated snow and meteorological variables, though coverage and variable selection vary widely, with few stations reporting full GCM-relevant parameters \cite{Raleigh_Livneh_Lapo_Lundquist_2015}. Example instruments for automated measurements include ultrasonic depth sensors, snow pillows, and tipping-bucket gauges, which often are noisy, downtime-prone, or yield unphysical values \cite{Hill_Burakowski_Crumley_Keon_Hu_Arendt_Wikstrom_Jones_Wolken_2019, Fleming_Zukiewicz_Strobel_Hofman_Goodbody_2023}. Some QC standards exist in some networks, such as SNOTEL (e.g., \citeNP{Serreze_Clark_Armstrong_McGinnis_Pulwarty_1999}), though many data streams remain raw. 

In prior work establishing our parameterization framework \cite{Charbonneau_Deck_Schneider_2025}, established quality control protocols were extended across additional variables and sites. Subsets of these procedures (denoted ``CQC" henceforth) were reused/modified in this work for snow albedo and SMM integration. MODIS and SNOTEL products generally agree once snow depth exceeds 5 cm \cite{Klein_2003}. While site slope/topography data are more rare, most SNOTEL stations are built on cleared flat ground, reducing the need for additional correction.

\subsection{Data Cleaning Methods}

For each of the 804 SNOTEL sites inside SMM bounds, from Oct 1 2000 to Oct 1 2023, daily \texttt{albedo\_dirty\_terrain\_corrected}, \texttt{snow\_fraction}, and \texttt{days\_without\_observation} fields from the corresponding SMM pixel (and 8 surrounding pixels) were paired with SNOTEL snow depth ($z$), snow water equivalent (SWE), air temperatures (daily average), and accumulated precipitation ($AP$). Site-years beginning collection after the start of the water-year (October 1st) or sites having under 60 SWE/precipitation values were excluded. SNOTEL values beyond instrument limits \cite{NEH622C2} were set to missing, and CQC steps 7–13 were applied. Zenith angle ($ZA$) at solar noon was calculated using \texttt{Insolations.jl} \cite{Insolation.jl} with UTC offsets derived from the free TimeZoneDB API (\url{https://timezonedb.com/api}), and precipitation was split into rain and snow using the bivariate logistics model from \cite{Jennings_Winchell_Livneh_Molotch_2018} ($>$85\% accuracy): 

\begin{linenomath*}
\begin{equation}
    f\ =\frac{1}{1\ +\ e^{\gamma+\beta T_a}}
\end{equation}
\end{linenomath*}

where $\beta = 1.24 ^\circ$C$^{-1}$ (temperature in Celsius) and $\gamma$ = -1.54. SNOTEL sites with radiation data used the same steps from June 1 2023 to June 1 2025, with the additional steps for hourly incoming ($SW_\mathrm{INC}$) and outgoing ($SW_\mathrm{OUT}$) shortwave radiation instead of the three SMM fields: 

\begin{enumerate}
	\item Set negative radiation values to zero 
	\item Derive hourly albedo $\alpha_i$ as $SW_\mathrm{OUT,i}$/ $SW_\mathrm{INC,i}$ 
	\item Excise all $\alpha_i$ outside [0-1] 
	\item Excise $\alpha_i$ with timestamp $ZA >$ 90$^\circ$ (using \texttt{Insolations.jl}) 
	\item Aggregate all $\alpha_i \leq 0.4$ to daily level (mean), and estimate $\alpha_g$ as the minimum of 0.1 or the median of all aggregated values having $z$ = 0 and $P$ = 0
	\item Aggregate all $\alpha_i >$ 0.35 to daily level (mean) to make $\alpha_\mathrm{snow}$ timeseries 
	\item Excise $\alpha_\mathrm{snow,i}$ where $z_i$ or SWE$_i$ are 0, missing, or SWE$_i > z_i$
	\item Excise $\alpha_\mathrm{snow,i}$ where $P_i$ = 0 but $d\mathrm{SWE}/dt_i > 0$ (avoid $d\alpha / dt$ samples under indeterminate snowfall) 
	\item Fill holes of 5 days or less in remaining $\alpha_\mathrm{snow}$ via linear interpolation
	\item Exclude sites with $<$ 60 remaining albedo values 
\end{enumerate}
For the SMM albedos (0-100), we used the following steps per site: 

\begin{enumerate}
    \item 	Excise \texttt{albedo\_dirty\_terrain\_corrected}$_i$ having \texttt{days\_without\_observation}$_i >$ 10, according to \citeA{RITTGER_Lenard_Palomaki_Brodzik_Stillinger_Bair_Dozier_Painter_2024} 
	\item Apply a power mean ($p$=3) over the remaining 3$\times$3 pixels of \texttt{albedo\_dirty\_terrain\_corrected} to mitigate geolocation error, extract the most representative ``snowy" estimate (mitigate non-snow or erroneous snow-fraction contamination, with a smoother non-max norm to avoid amplifying high-frequency noise or biasing towards possible cloud albedos, but more inclined to higher values than an arithmetic mean), and create a more complete $\alpha_\mathrm{snow}$ timeseries estimate. The count and standard deviation of utilized pixels was tracked. 
	\item Similarly apply a power mean ($p$= -3, now more inclined to lower values) over the same 3$\times$3 pixels of \texttt{albedo\_dirty\_terrain\_corrected}, and estimate $\alpha_g$ as the median of all means $\leq$ 40 where $z_i=\mathrm{SWE}_i = P_i = P_{i-1} = \texttt{days\_without\_observation}_i = 0$. Exclude sites lacking any samples for this estimate.
	\item Repeat steps 7-10 from above, excising $\alpha_\mathrm{snow,i} <$ 35 as ground-contaminated signal (this choice was corroborated by the standard deviation over utilized 3$\times$3 pixels at such values being $<$ 8, while snowy pixels exhibited abruptly higher variation) 
	\item Scale resulting albedo timeseries by 100 (to [0-1]) 

\end{enumerate}
Finally, all albedos with $z <$ 5 cm were excised to avoid small snowpack discrepancies.

\subsection{Pre-Training Procedures}

Remaining series exhibited unrealistic high-frequency oscillations, especially in-situ or during peak season, exceeding expected daily changes \cite{Sarangi_Qian_Rittger_Bormann_Liu_Wang_Wan_Lin_Painter_2019, Bair_Davis_Dozier_2018}. These could arise from transient shadows or wind-driven debris, local heterogeneity contamination from the above cleaning procedures, or systematic sensor issues requiring further consideration. At both coarse and site scales, persistent non-snow surfaces (rocks, vegetation, shadows, sensor arrays) can also skew true snow fraction even when reported snow cover fraction is 100\% at large snowpack, biasing reported albedo downward \cite{Etchevers_Martin_Brown_Fierz_Lejeune_Bazile_Boone_Dai_Essery_Fernandez_et_al._2004, Bair_Dozier_Stern_LeWinter_Rittger_Savagian_Stillinger_Davis_2022}. To address these issues, we applied a two-step correction:

\subsubsection{Exponential Smoothing}

An exponential moving average (EMA) was applied as a low-pass filter to dampen noise, extracting the trends relevant for climate modeling scales: 

\begin{linenomath*}
\begin{equation}
    EMA_i=w\alpha_i+\left(1-w\right)EMA_{i-1} 
\end{equation}
\end{linenomath*}

Other smoothing approaches exist (e.g. \citeNP{Ye_Cheng_Hao_Yu_Ma_Liang_Shen_2023, Calleja_CorbeaPerez_Fernandez_Recondo_Peon_de_Pablo_2019}, though we prefer the EMA due to simpler implementation, true low-pass filtering compared to techniques like max-filtering which can amplify noise, and better preservation of precipitation-driven albedo changes via weighting bias to recent samples. This approach will smooth out some more abrupt authentic transitions and influence the smoothness of the learned dynamics, but also aids in filling out the otherwise more discrete space of raw $d\alpha /dt$ targets due to data precision, which can throttle a neural network's learning ability. We chose $w = 0.33$  for SMM and $w =0.1$  for noisier in-situ data to balance noise suppression against preserving physically relevant trends. Each time the EMA was restarted due to missing/cleaned values, the first 5 values were discarded for sufficient burn-in.

\subsubsection{Rescaling}

A maximum-scaling was applied to correct for persistent non-snow contamination. Measured albedo is a mix of background ($\alpha_{bg}$) and snow signals weighted by snow fraction, $f$: 
\begin{linenomath*}
\begin{equation}
\alpha_{obs}=\left(1-f_{snow}\right)\alpha_{bg}\ +\ f_{snow}\alpha_{snow}
\end{equation}
\end{linenomath*}

If persistent non-snow contamination prevents full snow coverage ($f_\mathrm{max}<1$) even for peak snowpack in a state of maximum possible albedo ($\alpha_\mathrm{theory,max}$), the measured maximum albedo ($\alpha_\mathrm{obs,max}$) will be reduced:

\begin{linenomath*}
\begin{equation}
\alpha_\mathrm{obs,max}=\left(1-f_\mathrm{max}\right)\alpha_{bg}\ +\ f_\mathrm{max}\alpha_\mathrm{theory,max}
\end{equation}
\end{linenomath*}

Suitable choice of $\alpha_\mathrm{theory,max}$ and $\alpha_{bg}$ knowledge thus allows estimation of $f_\mathrm{max}$ and a scaling transformation to rescale albedos: 

\begin{linenomath*}
\begin{equation}
    f_\mathrm{max}=\frac{\alpha_\mathrm{obs,max}-\alpha_{bg}}{\alpha_\mathrm{theory,max}-\alpha_{bg}},\;\;\;\;    \alpha_i=\frac{\alpha_{obs,i}-\alpha_{bg}\left(1-f_\mathrm{max}\right)}{f_\mathrm{max}} 
\end{equation}
\end{linenomath*}

Although the best $\alpha_\mathrm{theory,max}$ (i.e. fresh pure snowfall on large snowpack) choice could fluctuate by site, it should exhibit tight range across climates and be well-represented within multi-year records, enabling estimation of $f_\mathrm{max}$. The top 5\% of all in-situ sites’ cleaned albedos were collected, using the mean of these values to estimate $\alpha_\mathrm{theory,max} = 0.88$, which aligns with theoretical maximums of broadband albedo \cite{Dozier_Green_Nolin_Painter_2009}. This rescaling serves to reduce non-snow contamination and potential unmeasured slope effects, and recorrects EMA-induced downward bias while also leaving series with $\alpha_\mathrm{obs,max}\approx\alpha_\mathrm{theory,max}$ largely unchanged. $\alpha_\mathrm{obs,max}$ was estimated with the 95th percentile of each site’s cleaned and smoothed albedo measurements, removing and interpolating any $\alpha_\mathrm{snow}>1$ after the transformation. 

Daily $d\alpha / dt$ was then computed as $\frac{\alpha_{i+1}-\alpha_i}{t_{i+1}-t_i}$, retaining only consecutive days (when
$t_{i+1}-t_i=1$ day). The final training set excluded cases with \texttt{snow\_fraction} $<$ 0.05 (the mean over the available 3$\times$3 pixels) and $z$/SWE = 1, though these were kept in testing data. Sites offering less than 30 training samples were excluded from the study.

For non-SNOTEL sites (excluding El Tapado and CUES), non-albedo cleaning is described in CQC, and albedo cleaning followed the steps for SNOTEL above. Col de Porte already gives daily albedo, requiring steps 5,7,8,9 only. For CUES, relative humidity was capped at 1, only albedo steps 4,6,8,9 were used, and snowfall was estimated using the Zwart parameterization \cite{zwart2007significance} from positive changes in $z$, in accordance with a site representative. No ground albedo was determined but $\alpha_\mathrm{obs,max}$ was sufficiently high to only apply the EMA. For El Tapado, the following steps were used: 

\begin{enumerate}
    \item Excise negative $z$, SWE, and $AP$; $AP > 1$ m, $T_a > 20^\circ$C (sensor noise) 
	\item Use only 2017 and 2019 water years from availability and quality concerns (per site correspondence) 
	\item Fill $T_a$ in 2017 via lapse-rate interpolation, in accordance with \cite{Voordendag_Reveillet_MacDonell_Lhermitte_2021}
	\item Calculate changes in $AP$, setting values $<$ 0.5 mm or negative to zero, then reconstruct cumulative AP from these differences. 
	\item Apply all SNOTEL albedo steps and CQC steps 9-13 
\end{enumerate}

The feature set was largely driven by near-term availability, computational simplicity, and quality considerations. Although rain can affect albedo \cite{Xie_Pettersen_Flanner_Shates_2024}, including $P_\mathrm{rain}$ had negligible impact on outcome (as in our prior study) and was omitted for computational simplicity. A 12-h offset exists between SNOTEL precipitation (local midnight) and SMM albedo (roughly local noon) measurements, which can blur interpretation of albedo changes and relationship ``separability". However, the EMA already coarse-grains changes and additional methods would likely further blur these relationships, so no further adjustment was applied.

\section{Performance Statistics}
\clearpage

\begin{sidewaystable}
\caption{\footnotesize \textbf{Parameterization scores over as-published parameterizations}. Performance of the different schemes in this study across both validation and testing sites, denoting models using their labels provided in Table 2 in the main text. Medians are listed, with the mean in parentheses alongside the median, for each of the performance metrics discussed in the text. For both coarse and point-level sites, each scheme also lists the p-value from the Wilcoxon signed-rank tests in this study, comparing each scheme to our proposed scheme ($NN$). Significant p-values ($<$ 0.05) are 
listed in bold, as well as the top scheme within each metric, unless no clear winner was determinable. Our scheme is one of the top performers for most metrics, and all statistical tests are significant ($p <$ 0.05) comparing the proposed scheme with regards to RMSE and RMSE\%. Scores are also similar between coarse and point-level sites for our model, indicating generalizability across spatial scales.}
\label{t:published-results}
\resizebox{\textwidth}{!}{%
\begin{tabular}{ccccc|cc|cccc|cc}
\hline
\multicolumn{1}{l}{} & \multicolumn{4}{c|}{Coarse Testing Data (500 m)} & \multicolumn{2}{c|}{Significance} & \multicolumn{4}{c|}{Site Testing Data (Point Observation)} & \multicolumn{2}{c}{Significance} \\
Model & RMSE\% & RMSE & Bias\% & Bias & RMSE\% & RMSE & RMSE\% & RMSE & Bias\% & Bias & RMSE\% & RMSE \\ \hline
$NN$ & \textbf{6.84 (7.29)} & \textbf{0.0494 (0.0509)} & 0.46 (0.57) & 0.0059 (0.0079) & - & - & \textbf{7.51 (8.05)} & \textbf{0.0555 (0.0584)} & 0.48 (-0.12) & 0.0076 (0.0038) & - & - \\
$CL$ & 14.83 (14.9) & 0.1095 (0.1089) & 2.93 (2.85) & 0.0282 (0.0288) & \textbf{2.87E-116} & \textbf{2.87E-116} & 14.03 (14.43) & 0.1050 (0.1061) & 2.15 (2.13) & 0.0199 (0.0223) & \textbf{9.31E-10} & \textbf{1.86E-09} \\
$OK$ & 13.02 (13.56) & 0.0929 (0.0959) & -0.26 (-0.02) & 0.0042 (0.0075) & \textbf{2.88E-116} & \textbf{2.96E-116} & 12.05 (13.3) & 0.0872 (0.0945) & -0.46 (0.17) & -0.0004 (0.0067) & \textbf{6.52E-09} & \textbf{1.3E-8} \\
$LV$ & 24.64 (24.84) & 0.1905 (0.1918) & 18.27 (18.04) & 0.1433 (0.1429) & \textbf{2.87E-116} & \textbf{2.87E-116} & 22.28 (23.82) & 0.1706 (0.1843) & 15.06 (15.73) & 0.1192 (0.1259) & \textbf{9.31E-10} & \textbf{9.31E-10} \\
$MD$ & 12.27 (12.88) & 0.0858 (0.0870) & -2.4 (-2.71) & -0.0109 (-0.0126) & \textbf{6.06E-116} & \textbf{1.02E-115} & 11.67 (12.91) & 0.0855 (0.0887) & -2.11 (-2.84) & -0.0104 (-0.0144) & \textbf{9.31E-10} & \textbf{1.86E-09} \\
$JL$ & 11.06 (11.59) & 0.0732 (0.0741) & -4.44 (-4.68) & -0.0276 (-0.0283) & \textbf{1.3E-111} & \textbf{1.51E-104} & 10.56 (12.15) & 0.0723 (0.0800) & -4.77 (-5.99) & -0.0311 (-0.0376) & \textbf{6.52E-08} & \textbf{4.0E-6} \\
$SP$ & 29.29 (29.91) & 0.2082 (0.2157) & 1.68 (2.97) & 0.0046 (0.015) & \textbf{2.87E-116} & \textbf{2.87E-116} & 29.26 (29.85) & 0.2123 (0.2167) & -8.52 (-6.59) & -0.0761 (-0.0509) & \textbf{9.31E-10} & \textbf{9.31E-10} \\
$MS$ & 14.21 (14.25) & 0.1097 (0.1093) & 7.39 (7.11) & 0.0634 (0.0612) & \textbf{2.87E-116} & \textbf{2.87E-116} & 13.5 (14.11) & 0.1041 (0.1081) & 6.61 (5.68) & 0.0564 (0.0503) & \textbf{9.31E-10} & \textbf{9.31E-10} \\
$UT$ & 17.14 (17.35) & 0.1261 (0.1294) & 4.95 (5.24) & 0.0435 (0.0462) & \textbf{2.87E-116} & \textbf{2.87E-116} & 16.39 (17.02) & 0.1183 (0.1256) & 3.28 (4.62) & 0.0306 (0.0398) & \textbf{9.31E-10} & \textbf{9.31E-10} \\
$HT$ & 13.57 (13.92) & 0.0986 (0.1011) & 3.7 (3.96) & 0.0309 (0.0339) & \textbf{2.87E-116} & \textbf{2.87E-116} & 13.94 (13.76) & 0.0985 (0.1011) & 2.27 (2.39) & 0.0205 (0.022) & \textbf{9.31E-10} & \textbf{2.79E-09} \\
$EC$ & 15.98 (16.32) & 0.1147 (0.1157) & -1.71 (-1.74) & -0.0105 (-0.0089) & \textbf{2.87E-116} & \textbf{2.87E-116} & 15.71 (16.4) & 0.1154 (0.1163) & -3.2 (-3.33) & -0.0227 (-0.0204) & \textbf{9.31E-10} & \textbf{1.86E-09} \\
$ZO$ & 28.9 (29.91) & 0.2267 (0.2329) & 27.18 (27.91) & 0.2138 (0.2178) & \textbf{2.87E-116} & \textbf{2.87E-116} & 29.29 (29.45) & 0.2348 (0.2301) & 26.75 (25.9) & 0.2113 (0.2045) & \textbf{9.31E-10} & \textbf{9.31E-10} \\
$IT$ & 39.17 (39.09) & 0.3059 (0.3063) & 37.87 (37.83) & 0.2946 (0.2948) & \textbf{2.87E-116} & \textbf{2.87E-116} & 35.74 (36.26) & 0.2816 (0.2851) & 33.02 (34.28) & 0.2582 (0.2673) & \textbf{9.31E-10} & \textbf{9.31E-10} \\
$ID$ & 13.2 (13.56) & 0.0859 (0.0865) & -7.99 (-8.0) & -0.0545 (-0.0534) & \textbf{4.46E-115} & \textbf{1.02E-112} & 12.19 (14.22) & 0.0874 (0.0939) & -8.33 (-8.87) & -0.06 (-0.0594) & \textbf{1.86E-09} & \textbf{1.77E-08} \\
$BA$ & 11.32 (11.69) & 0.0778 (0.0795) & -0.24 (-0.29) & \textbf{0.0029 (0.0037)} & \textbf{2.87E-116} & \textbf{2.88E-116} & 11.43 (11.72) & 0.0813 (0.0810) & -1.42 (-1.77) & -0.0058 (-0.007) & \textbf{1.3E-8} & \textbf{2.86E-7} \\
$VA$ & 21.55 (23.13) & 0.1588 (0.1740) & 4.46 (6.62) & 0.0371 (0.0555) & \textbf{2.87E-116} & \textbf{2.87E-116} & 24.28 (26.5) & 0.1776 (0.1900) & 4.1 (6.83) & 0.0321 (0.0525) & \textbf{9.31E-10} & \textbf{9.31E-10} \\
$LE$ & 15.31 (16.07) & 0.1047 (0.1101) & 2.44 (2.62) & 0.0228 (0.0245) & \textbf{2.87E-116} & \textbf{2.87E-116} & 12.4 (13.27) & 0.0846 (0.0898) & \textbf{0.32 (0.06)} & 0.0073 (0.0054) & \textbf{6.52E-08} & \textbf{2.36E-7} \\
$AC$ & 28.02 (27.89) & 0.2192 (0.2175) & 20.22 (19.78) & 0.1612 (0.1577) & \textbf{2.87E-116} & \textbf{2.87E-116} & 26.09 (27.44) & 0.2040 (0.2124) & 17.16 (19.07) & 0.1357 (0.1507) & \textbf{9.31E-10} & \textbf{9.31E-10} \\
$GC$ & 16.4 (16.67) & 0.1065 (0.1065) & -10.73 (-10.88) & -0.0734 (-0.0737) & \textbf{3.88E-116} & \textbf{7.07E-116} & 16.22 (16.68) & 0.1062 (0.1090) & -10.97 (-11.43) & -0.0746 (-0.0776) & \textbf{1.86E-09} & \textbf{2.79E-09} \\
$NG$ & 15.84 (15.8) & 0.1178 (0.1181) & 5.31 (5.24) & 0.0464 (0.0464) & \textbf{2.87E-116} & \textbf{2.87E-116} & 15.46 (15.59) & 0.1130 (0.1169) & 4.25 (4.96) & 0.0375 (0.0434) & \textbf{9.31E-10} & \textbf{9.31E-10} \\
$ED$ & 27.88 (27.77) & 0.2167 (0.2159) & 18.42 (17.93) & 0.1464 (0.1433) & \textbf{2.87E-116} & \textbf{2.87E-116} & 26.3 (27.38) & 0.2022 (0.2111) & 15.35 (17.08) & 0.1201 (0.1353) & \textbf{9.31E-10} & \textbf{9.31E-10} \\
$ET$ & 23.42 (23.43) & 0.1722 (0.1748) & 12.04 (12.47) & 0.0935 (0.0972) & \textbf{2.87E-116} & \textbf{2.87E-116} & 21.24 (22.2) & 0.1625 (0.1656) & 7.66 (9.57) & 0.0643 (0.0759) & \textbf{9.31E-10} & \textbf{9.31E-10} \\
$BR$ & 25.78 (26.08) & 0.2028 (0.2054) & 22.91 (23.06) & 0.1808 (0.1825) & \textbf{2.87E-116} & \textbf{2.87E-116} & 24.4 (25.07) & 0.1939 (0.1961) & 21.86 (21.97) & 0.1741 (0.1724) & \textbf{1.86E-09} & \textbf{9.31E-10} \\
$IS$ & 11.46 (11.91) & 0.0756 (0.0773) & -4.24 (-4.35) & -0.0255 (-0.0255) & \textbf{4.02E-113} & \textbf{4.2E-110} & 10.73 (11.99) & 0.0751 (0.0796) & -4.38 (-5.24) & -0.028 (-0.0322) & \textbf{1.57E-07} & \textbf{7.01E-6} \\ \hline
\end{tabular}%
}
\end{sidewaystable}
\clearpage

\begin{sidewaystable}
\caption{\footnotesize \textbf{Parameterization scores over calibrated parameterizations}. Labeling and formatting follows that as described in Table \ref{t:published-results}. In this case, the proposed scheme is still  dominant in many metrics. Results of statistical tests are again all significant ($p <$ 0.05) with the exceptions of $HT$ over point-level data.}
\label{t:calibrated-results}
\resizebox{\textwidth}{!}{%
\begin{tabular}{ccccc|cc|cccc|cc}
\hline
\multicolumn{1}{l}{} & \multicolumn{4}{c|}{Coarse Testing Data (500 m)} & \multicolumn{2}{c|}{Significance} & \multicolumn{4}{c|}{Site Testing Data (Point Observation)} & \multicolumn{2}{c}{Significance} \\
Model & RMSE\% & RMSE & Bias\% & Bias & RMSE\% & RMSE & RMSE\% & RMSE & Bias\% & Bias & RMSE\% & RMSE \\ \hline
$NN$ & \textbf{6.84 (7.29)} & \textbf{0.0494 (0.0509)} & 0.46 (0.57) & 0.0059 (0.0079) & - & - & \textbf{7.51 (8.05)} & \textbf{0.0555 (0.0584)} & 0.48 (-0.12) & 0.0076 (0.0038) & - & - \\
$CL$ & 9.62 (10.1) & 0.0674 (0.0680) & -1.14 (-1.23) & -0.0029 (-0.0024) & \textbf{8.03E-109} & \textbf{4.46E-104} & 8.8 (9.83) & 0.0646 (0.0677) & -0.7 (-1.34) & -0.0017 (-0.0031) & \textbf{3.15E-04} & \textbf{4.44E-03} \\
$OK$ & 10.84 (11.52) & 0.0750 (0.0770) & -0.94 (-1.24) & \textbf{0.0003 (-0.0005)} & \textbf{1.46E-114} & \textbf{7.17E-113} & 10.14 (11.17) & 0.0723 (0.0770) & -0.77 (-1.62) & 0.001 (-0.0039) & \textbf{1.93E-07} & \textbf{4.99E-07} \\
$LV$ & 10.21 (10.85) & 0.0720 (0.0737) & -0.28 (-0.41) & 0.0045 (0.0049) & \textbf{1.51E-114} & \textbf{1.26E-112} & 9.91 (11.47) & 0.0758 (0.0792) & -0.68 (-1.77) & 0.0001 (-0.0049) & \textbf{4.00E-08} & \textbf{1.57E-07} \\
$MD$ & 10.96 (11.55) & 0.0760 (0.0777) & -0.46 (-0.79) & 0.0037 (0.0031) & \textbf{4.68E-115} & \textbf{6.86E-114} & 10.13 (11.43) & 0.0722 (0.0790) & -0.2 (-1.11) & 0.005 (-0.0001) & \textbf{5.12E-08} & \textbf{1.02E-07} \\
$JL$ & 8.65 (9.26) & 0.0639 (0.0653) & 1.65 (1.57) & 0.0171 (0.0179) & \textbf{2.32E-109} & \textbf{2.39E-112} & 8.26 (9.43) & 0.0639 (0.0670) & -0.11 (-0.44) & 0.0064 (0.0032) & \textbf{1.04E-02} & \textbf{9.15E-03} \\
$SP$ & 8.49 (9.01) & 0.0596 (0.0609) & -0.02 (0.08) & 0.0047 (0.0062) & \textbf{3.66E-107} & \textbf{1.43E-96} & 9.7 (10.56) & 0.0673 (0.0732) & -1.99 (-2.49) & -0.012 (-0.0122) & \textbf{1.19E-05} & \textbf{2.82E-05} \\
$MS$ & 8.36 (9.11) & 0.0591 (0.0612) & -0.29 (-0.44) & 0.0023 (0.0022) & \textbf{6.45E-107} & \textbf{1.63E-92} & 11.11 (11.89) & 0.0815 (0.0831) & -3.22 (-2.77) & -0.0167 (-0.0138) & \textbf{1.28E-07} & \textbf{4.16E-07} \\
$UT$ & 9.71 (10.16) & 0.0682 (0.0691) & -0.99 (-1.01) & -0.0013 (-0.001) & \textbf{1.38E-111} & \textbf{4.58E-109} & 8.94 (10.11) & 0.0635 (0.0707) & 0.05 (-0.58) & 0.0034 (0.0025) & \textbf{1.18E-04} & \textbf{1.72E-03} \\
$HT$ & 8.26 (8.54) & 0.0599 (0.0617) & 1.63 (1.89) & 0.016 (0.0192) & \textbf{1.18E-87} & \textbf{1.79E-98} & 7.97 (8.5) & 0.0633 (0.0638) & 3.36 (2.06) & 0.0283 (0.0209) & 2.39E-01 & 6.64E-02 \\
$EC$ & 9.72 (10.43) & 0.0680 (0.0703) & -0.35 (-0.46) & 0.0038 (0.0036) & \textbf{2.34E-115} & \textbf{1.25E-114} & 9.71 (10.9) & 0.0692 (0.0750) & -1.18 (-2.14) & -0.0043 (-0.0086) & \textbf{1.57E-07} & \textbf{2.57E-06} \\
$ZO$ & 9.78 (10.26) & 0.0666 (0.0686) & -0.77 (-0.62) & 0.0008 (0.0025) & \textbf{8.1E-114} & \textbf{6.06E-110} & 10.46 (11.7) & 0.0762 (0.0810) & -0.73 (-1.79) & -0.0003 (-0.0054) & \textbf{4.99E-07} & \textbf{3.46E-06} \\
$IT$ & 9.25 (9.81) & 0.0657 (0.0666) & 0.03 (-0.01) & 0.0058 (0.0066) & \textbf{7.01E-115} & \textbf{2.82E-113} & 9.52 (10.96) & 0.0693 (0.0762) & -2.06 (-2.33) & -0.0096 (-0.0102) & \textbf{2.86E-07} & \textbf{2.57E-06} \\
$ID$ & 9.53 (9.99) & 0.0648 (0.0668) & -0.7 (-0.57) & 0.0003 (0.0024) & \textbf{2.99E-112} & \textbf{1.09E-106} & 10.05 (11.26) & 0.0734 (0.0783) & -1.12 (-1.93) & -0.0012 (-0.0067) & \textbf{1.38E-06} & \textbf{1.35E-05} \\
$BA$ & 9.28 (9.92) & 0.0656 (0.0674) & -0.28 (-0.41) & 0.0029 (0.0032) & \textbf{2.53E-112} & \textbf{8.42E-113} & 8.8 (9.95) & 0.0617 (0.0684) & -1.29 (-2.31) & -0.0041 (-0.011) & \textbf{3.15E-04} & \textbf{1.47E-03} \\
$VA$ & 10.19 (10.72) & 0.0754 (0.0768) & 2.67 (2.47) & 0.0276 (0.027) & \textbf{4.32E-116} & \textbf{4.33E-116} & 9.75 (10.83) & 0.0725 (0.0782) & 1.92 (1.19) & 0.0214 (0.0173) & \textbf{2.21E-06} & \textbf{3.46E-06} \\
$LE$ & 10.02 (10.62) & 0.0693 (0.0714) & -0.47 (-0.57) & 0.0028 (0.0028) & \textbf{1.36E-115} & \textbf{6.25E-115} & 9.19 (10.83) & 0.0655 (0.0735) & -1.47 (-2.33) & -0.0056 (-0.01) & \textbf{1.17E-06} & \textbf{3.57E-05} \\
$AC$ & 10.97 (11.64) & 0.0782 (0.0796) & 0.2 (-0.12) & 0.0093 (0.0081) & \textbf{1.37E-115} & \textbf{5.74E-115} & 10.14 (11.21) & 0.0732 (0.0781) & 0.13 (-0.88) & 0.0081 (0.002) & \textbf{1.93E-07} & \textbf{5.96E-07} \\
$GC$ & 11.18 (11.75) & 0.0778 (0.0801) & 0.01 (-0.13) & 0.0091 (0.0086) & \textbf{7.01E-116} & \textbf{2.09E-115} & 10.26 (11.33) & 0.0737 (0.0795) & 0.54 (-0.48) & 0.008 (0.0054) & \textbf{8.42E-07} & \textbf{1.90E-06} \\
$NG$ & 10.65 (11.27) & 0.0755 (0.0768) & 0.09 (-0.27) & 0.0084 (0.0067) & \textbf{1.12E-114} & \textbf{1.11E-113} & 9.52 (10.66) & 0.0693 (0.0738) & 0.29 (-0.89) & 0.008 (0.0016) & \textbf{2.99E-06} & \textbf{8.02E-06} \\
$ED$ & 10.78 (11.39) & 0.0776 (0.0789) & 1.03 (0.73) & 0.0156 (0.0147) & \textbf{2.16E-115} & \textbf{5.09E-115} & 9.79 (10.92) & 0.0722 (0.0771) & 1.02 (0.09) & 0.0139 (0.0095) & \textbf{7.10E-07} & \textbf{1.17E-06} \\
$ET$ & 9.37 (10.0) & 0.0648 (0.0669) & -0.46 (-0.55) & 0.003 (0.0027) & \textbf{8.56E-114} & \textbf{5.1E-111} & 8.92 (10.5) & 0.0639 (0.0716) & -1.55 (-2.34) & -0.0057 (-0.0103) & \textbf{9.16E-06} & \textbf{2.15E-04} \\
$BR$ & 10.17 (10.81) & 0.0730 (0.0741) & 0.58 (0.33) & 0.0115 (0.0107) & \textbf{1.77E-115} & \textbf{5.55E-115} & 8.9 (10.27) & 0.0662 (0.0727) & 0.97 (-0.35) & 0.0134 (0.0048) & \textbf{4.00E-06} & \textbf{6.11E-06} \\
$IS$ & 8.92 (9.62) & 0.0634 (0.0649) & -0.33 (-0.4) & 0.0026 (0.0032) & \textbf{4.54E-109} & \textbf{1.95E-108} & 8.56 (10.07) & 0.0641 (0.0694) & -1.86 (-2.39) & -0.0077 (-0.0114) & \textbf{2.15E-04} & \textbf{1.59E-03} \\ \hline
\end{tabular}%
}
\end{sidewaystable}
\clearpage

\section{Interaction Strength}\label{a:interaction}

\begin{table*}[ht!]
\caption{\footnotesize \textbf{Feature Interaction Strength}. This table displays the pairwise Friedman H-Statistics of each feature pair in the neural network. No value exceeds 0.2, implying minimal higher-order effects compared to the additive effects diagnosed from Fig. \ref{f:ale}.}
\label{t:interaction}
\begin{center}
\begin{tabular}{cc}
\hline
\multicolumn{2}{c}{H-Statistic} \\ \hline
\multicolumn{1}{c|}{$P$, $\mu$} & 0.13 \\
\multicolumn{1}{c|}{$P$, $\alpha$} & 0.11 \\
\multicolumn{1}{c|}{$P$, $T_a$} & 0.12 \\
\multicolumn{1}{c|}{$\mu$, $T_a$} & 0.18 \\
\multicolumn{1}{c|}{$\alpha$, $T_a$} & 0.04 \\
\multicolumn{1}{c|}{$\alpha$, $\mu$} & 0.09 \\ \hline
\end{tabular}%
\end{center}
\end{table*}

The pairwise Friedman H-statistic \cite{molnar2025} quantifies the strength of an interaction between two model features. The square of the statistic can be interpreted as quantifying what fraction of the prediction's variance comes from solely the interaction effect between two features rather than their additive main effects. For this network, all values are less than 0.2, implying weak pairwise interactions (less than 5\% of the prediction variance stems from interaction effects). This means the overall prediction effects can be summarized as primarily being additive in the main single-variable effects (those assessed in the ALE in Fig. \ref{f:ale}), instead of also including strong higher-order effects. The reported statistics remain stable between different binning sizes as well as under consideration of the feature correlations, implying the results are faithful for the developed scheme and robust against numerical artifacts that can arise for a particular calculation approach.

\section{Hyperparameters}\label{a:hyper}
Similar to our previous work \cite{Charbonneau_Deck_Schneider_2025}, we made use of a custom loss-function:
\begin{equation} 
    L = \frac{1}{N_d} \sum_{i=1}^{N_d}  w_i |y_i-\hat{y}_i |^{n_1}\;,
    \;\;\;\; w_i = 1 + |y_i|^{n_2}.
\end{equation}
where $N_d$ is the number of batched training data, $\hat{y}_i$ and $y_i$ are the (scaled) prediction and target data, and $n_1$ and $n_2$ are constant positive numbers. Using $(n_1 = 1$, $n_2 = 0$) or ($n_1 = 2$, $n_2 = 0$) is equivalent to optimizing the average $L_1$ or $L_2$ losses, respectively. This function enables investigation of emphasizing extremes, which can be important in prognostic (accumulated) modeling as undershooting one extreme $d\alpha/dt$ can propagate the offset to future time steps. Optimal hyperparameters and the assessed values are summarized in Table C1, for $n_1$, $n_2$, and layer width hyperparameter $n$. Scores for each configuration were evaluated using leave-one-out validation (training on 39 of the 40 sites, tracking the timeseries RMSE\% and bias (B) of the remaining site, and averaging over the 40 possible combinations) with a batch size of 256, tracking performance every 10 epochs over 200 epochs. Most timeseries trials performed optimally when training for approximately 100 epochs with the small network, similar to our previous findings.

\begin{table*}[ht!]
\caption{\footnotesize \textbf{Hyperparameter results}. Timeseries RMSE\% and B\% were used as the primary metrics. $n_1$ = 2 ($L_2$-like metric) provided the lowest RMSE\% for all choices.}
\label{t:hyperparam}
\begin{center}
\begin{tabular}{cccc}
\hline
Parameter & Description & Range & Best Value\\ \hline
$n$ & Width of Mixing Layer & 2, 3, 4, 5 & 2\\
$n_1$ & Power Scaling of Prediction Error & 1, 2, 3& 2\\
$n_2$ & Power Scaling of Target Magnitude & 0, 0.1, 1, 2 & 0\\
\hline
\end{tabular}
\end{center}
\end{table*}

The configuration of $n=2$, $n_1 = 2$, and $n_2 = 0$ showed optimal performance for both timeseries RMSE\% and B. As $n_2 = 0$ matches the $L_2$ loss function, true $L_2$ was used for final model training. We note $n_2 \neq 0$ and/or $n_1 = 1$ did often exhibit timeseries that better replicated sub-weekly variation, but at the expense of whole-timeseries offsets that skewed overall performance statistics (which would similarly skew the energy balance within a GCM) to favor $n_2 = 0$. As $L_2$ loss is associated with RMSE, it is less surprising that $n_1 = 2$ exhibited optimal RMSE\%, though optimal $n_2 = 0$ differs from our previous study. This optimal configuration is likely influenced by the relative lack of variation in the coarse-scale (and cleaned) albedo data used for training, compared to the testing site-level data with more variation or the site-level snow depth data in our previous work.

%



%
%

\section*{Open Research Section}
All utilized data are publicly available. The SNOTEL data utilized in this study are available via the National Water and Climate Center. SNOTEL Data reports were generated using the online portal found at \url{https://www.nrcs.usda.gov/wps/portal/wcc/home/}. Links to raw and processed data for Col De Porte \cite{Lejeune_Dumont_Panel_Lafaysse_Lapalus_Le_Gac_Lesaffre_Morin_2019}, K{\"u}htai \cite{Krajci_Kirnbauer_Parajka_Schoer_Bloschl_2017}, Sodankyla \cite{Essery_Kontu_Lemmetyinen_Dumont_Menard_2016}, Yala \cite{Stigter_Steiner_Koch_Saloranta_Kirkham_Immerzeel_2021, Shea_Wagnon_Immerzeel_Biron_Brun_Pellicciotti_2015}, and Upper Rofental \cite{Warscher_Marke_Rottler_Strasser_2024} data can be found in our previous work \cite{Charbonneau_Deck_Schneider_2025}. SMM data can be found at the Snow Today website, hosted by the National Snow and Ice Data Center (NSIDC) \url{https://nsidc.org/data/stc_modscgdrf_hist/versions/1} \cite{RITTGER_Lenard_Palomaki_Brodzik_Stillinger_Bair_Dozier_Painter_2024}. El Tapado raw data was retrieved from the CEAZA meteorological network data portal at \url{https://www.ceazamet.cl/index.php?e_cod=TPF&pag=mod_estacion}, and CUES data can be found for download at \url{https://snow.ucsb.edu/index.php/level-2-model-ready} \cite{Bair_Davis_Dozier_2018}. The \texttt{Insolations.jl} package is developed by the Climate Modeling Alliance (CliMA), able to be downloaded at \url{https://clima.github.io/Insolation.jl/dev/}  \cite{Insolation.jl}, while the API for time-zone retrieval can be found at \url{https://timezonedb.com/api}. As in our previous work, code for scraping and cleaning data, CSVs of data, and tutorials for data retrieval and modifying our (or building custom) constrained neural networks are available at \url{https://clima.github.io/ClimaLand.jl/dev/generated/standalone/Snow/base_tutorial/}, to provide quality-controlled datasets for calibration and ML applications. All data cleaning procedures are described above in Appendix A. Time-stepping is described in the main text. Descriptions and equations of the compared models are provided in the Supporting Information.

\section*{Conflict of Interest disclosure}
The authors declare there are no conflicts of interest for this manuscript.

\acknowledgments
We thank Edward Bair and Annelies Voordendag for their correspondence regarding CUES and El Tapado data, respectively, and for Cristian Nelson for correspondence about El Tapado data. We thank the Col de Porte, Kuhtai, Sodankyla, Upper Rofental, CUES, CEAZA, and Yala Basecamp teams for their data. 

A. C. was supported by the AI4Science initiative at the California Institute for Technology and  a Department of Defense National Defense Science and Engineering Graduate (NDSEG) Fellowship.
All authors were generously supported by Schmidt Sciences, L.L.C. and by the Resnick Sustainability Institute.

\textbf{Author contributions}: \\
Conceptualization: AC\\
Methodology: AC, KD, TS\\
Investigation: AC\\
Visualization: AC\\
Supervision: KS, TS \\
Writing – original draft: AC \\
Writing – review: KD, TS \\

%
%

\bibliography{main.bib}

\nociteSupp{*}
\bibliographystyleSupp{apacite}
\bibliographySupp{si.bib}

%
%
%
%
%

\end{document}